\documentclass[reprint,twocolumn,prb,amsmath,amssymb,aps]{revtex4-2} 

\usepackage{graphicx}  
\usepackage{dcolumn}   
\usepackage{bm}        
\usepackage{hyperref}  
\hypersetup{colorlinks=true, linkcolor=blue!95!black!85!yellow, citecolor=blue!95!black!85!yellow, urlcolor=blue!95!black!85!yellow} 
\usepackage{physics}   
\usepackage{newtxtext} 
\usepackage[cmintegrals]{newtxmath} 
\usepackage{booktabs}  
\usepackage{multirow}  
\usepackage{xcolor}    
\usepackage{float}     

\usepackage{tikz}
\newcommand{\orcidicon}{%
    \begin{tikzpicture}
        \draw[lime, fill=lime] (0,0) circle [radius=0.16] node[white] {{\fontfamily{qag}\selectfont \tiny ID}};
        \draw[white, fill=white] (-0.0625,0.006) circle [radius=0.007];
    \end{tikzpicture}
    \hspace{-2.9mm}
}
\foreach \x in {A, B}{\expandafter\xdef\csname orcid\x\endcsname{\noexpand\href{https://orcid.org/\csname orcidauthor\x\endcsname}{\noexpand\orcidicon}}}

\newcommand{\iitm}{School of Mechanical and Materials Engineering, Indian Institute of Technology Mandi, Kamand 175075, India}

\newcommand{\cm}{cm$^{-3}$}
\newcommand{\cvs}{cm$^{2}$ V$^{-1}$ s$^{-1}$}
\newcommand{\sbku}{$\mu$V/K}

\begin{document}

\preprint{APS/123-QED}

 \title{Band gap renormalization, carrier mobility, and transport in Mg$_{2}$Si and Ca$_{2}$Si:\\ \textit{Ab initio} scattering and Boltzmann transport equation study}

\author{Vinod Kumar Solet \orcidA{}}
\email{vsolet5@gmail.com}
\author{Sudhir K. Pandey \orcidB{}}
\email{sudhir@iitmandi.ac.in}
\affiliation{\iitm}
	
\begin{abstract}
 
We perform first-principles electron-phonon interaction (EPI) calculations based on many-body perturbation theory to study the temperature-dependent band-gap and charge-carrier transport properties for Mg$_{2}$Si and Ca$_{2}$Si using the Boltzmann transport equation (BTE) under different relaxation-time approximations (RTAs). For a PBE band gap of $\sim$0.21 (0.56) eV in Mg$_{2}$Si (Ca$_{2}$Si), a zero-point renormalization correction of $\sim$29-33 (37-51) meV is obtained using various approaches, while the gap at 300 K is $\sim$0.15-0.154 (0.46-0.5) eV. The electron mobility ($\mu_{\mathrm{e}}$), with a detailed convergence study at 300 K, is evaluated using linearized (self-energy and momentum RTA, or SERTA and MRTA) and iterative BTE (IBTE) solutions. At 300 K, the $\mu_{\mathrm{e}}$ values are $\sim$351 (100), 573 (197), and 524 (163) {\cvs} from SERTA, MRTA, and IBTE, respectively, for Mg$_{2}$Si (Ca$_{2}$Si). These respective values at 900 K decrease to $\sim$33 (11), 48 (21), and 28 (11) {\cvs}. SERTA (MRTA) provides results in better agreement with IBTE at higher (lower) temperatures, while SERTA-derived $\mu_{\mathrm{e}}$ closely matches experimental $\mu_{\mathrm{e}}$ values for Mg$_{2}$Si. All electrical parameters related to thermoelectrics are significantly influenced by the choice of RTA, with SERTA and MRTA yielding improved agreement with experimental results compared to constant RTA (CRTA) for Mg$_{2}$Si. Finally, we obtain the figure of merit under various modeling conditions, in which several promising strategies such as nanostructuring (phonon-boundary scattering) and mass-difference scattering are employed to suppress the lattice thermal conductivity computed within phonon-phonon interactions. This study clearly identifies the critical role of different scattering mechanisms (mainly EPI) in accurate transport predictions of thermoelectric silicides.

\end{abstract}

\maketitle
\section{Introduction} 
\setlength{\parindent}{3em}

Understanding the dynamics of charge and energy transport processes in semiconductors has long been a topic of intense interest, encompassing both applied and fundamental aspects of condensed matter and materials physics \cite{ashcroft1976,ziman1960}. These transport phenomena are, in fact, crucial to technological advancements, as they govern the performance of solar cells, thermoelectric (TE) generators, batteries, and other relevant devices \cite{kirchartz2018makes,wood1988materials,park2010review}. Among these technologies, TE devices are particularly noteworthy for their ability to convert heat energy into electricity via the Seebeck effect \cite{snyder2008complex,shastri2021theory}. The TE energy conversion efficiency is typically quantified by the figure of merit (\textit{z}T), which is fundamentally proportional to the square of the Seebeck coefficient (S), electrical conductivity ($\sigma$) and absolute temperature (T), while inversely proportional to the thermal conductivity ($\kappa_{\mathrm{e}}$+$\kappa_{\mathrm{ph}}$), where $\kappa_{\mathrm{e}}$ and $\kappa_{\mathrm{ph}}$ are the contribution from the electrical carriers (both electrons and holes) and from the quantized lattice vibrations (phonons), respectively \cite{ayachi2023solar}.  

A primary hurdle in TE power conversion is achieving a high \textit{z}T (greater than 1), primarily due to the interdependent and competing nature of the charged carrier transport parameters. For example, at high temperatures, $\sigma$ decreases because of increased carrier scattering, while S decreases when carrier concentrations become too high, making it difficult to achieve an optimal balance. Meanwhile, $\kappa_{\mathrm{e}}$ increases with concentration, further complicating efforts to optimize $z$T. One effective strategy for enhancing $z$T involves boosting the product $\sigma$S$^{2}$ and decreasing $\kappa_{\mathrm{ph}}$ by adjusting the doping concentration \cite{liu2012convergence,sofo1994optimum}. Over the last few years, significant progress has also been made in reducing $\kappa_{\mathrm{ph}}$ through enhanced phonon scattering achieved by nanostructuring \cite{dresselhaus2007new,poudel2008high,li2012thermal,hochbaum2008enhanced}.

TE materials are typically composed of highly doped narrow band gap semiconductors \cite{sofo1994optimum}. Developing advanced materials for energy applications remains a central challenge in materials science. It requires a clear understanding of the physics underlying transport properties to improve existing materials and discover new ones. To effectively screen potential TE materials, transport coefficients must either be theoretically predicted or experimentally measured. Meanwhile, computational techniques with strong predictive capabilities \cite{curtarolo2013high,alberi20182019,li2024deep} offer distinct advantages over experiments. Predicting materials via first-principles often rely on electronic structure and phonon dispersion, but estimating transport properties like mobility and thermal conductivity requires an understanding of how particle (electron or phonon) evolve under the influence of other states \cite{ponce2020first,bernardi2016first}. This is achieved by combining the Boltzmann transport equation (BTE) with \textit{ab initio} methods \cite{zhou2016first,ponce2020first,bernardi2016first}.

A significant challenge in solving the Boltzmann equation for either charge carrier or phonon transport is understanding particle scattering caused by other particles (or by defects and boundaries). This critical information is not captured by density functional theory (DFT) due to its non-interacting framework used for deriving the energy band diagram. As a result, the constant relaxation time approximation (CRTA) is typically applied to estimate charge carrier transport coefficients \cite{singh2016investigation,shastri2020thermoelectric,solet2022first,sihmar2024understanding}. The CRTA is a simple and useful method in certain cases, such as for metals, but it can be less effective for semiconductors due to the non-ignorable dependence of charge carrier lifetimes on dispersion \cite{ashcroft1976,bernardi2014ab}. In thermal transport, for $\kappa_{\mathrm{ph}}$, phonons lifetimes are similarly influenced by interactions with other phonons, charge carriers, and defects etc. \cite{wen2020large,liao2015significant}. These complexities limit fully first-principles calculations and demand methodologies that extend beyond DFT to achieve accurate transport predictions.
 
For TE as a whole, it is essential to consider both particles (electrons and phonons), along with their coupling \cite{zhou2016first}. Consequently, electron–phonon (e-p) interactions (EPI) can play a crucial role in semiconductors, as they strongly affect scattering phenomena \cite{ziman1960,giustino2017electron}. In typical semiconductors, one of the causes of temperature-dependent features is the thermal motion of atoms within the crystal lattice, which affects the electronic states; a phenomenon known as e-p renormalization (EPR) \cite{giustino2010electron}. In calculations, it is common to combine DFT and many-body perturbation theory (MBPT) to address EPR \cite{ponce2015temperature}. Within this framework, the associated e-p scattering is typically assumed to be weak enough to approximate it as a first-order response to nuclear displacements, referred to as e-p coupling elements. In turn, MBPT then provides the renormalized charged electronic excitations, i.e., electron self-energy (SE). The real part of SE describes the band-structure renormalization, including both quantum zero-point renormalization (ZPR) and temperature effects \cite{gonze2011theoretical,cannuccia2012zero,antonius2015dynamical,capaz2005temperature,marini2008ab,cannuccia2011effect,antonius2014many}. This topic can be particularly important for achieving near-quantitative agreement between \textit{ab initio} results and experimental data \cite{querales2019temperature,zhang2020temperature,huang2021antibonding}. The imaginary part of SE corresponds to the e-p scattering cross section and is typically used to compute phonon-limited mobility via the BTE \cite{brunin2020phonon,ponce2020first}.

Solving the exact or iterative BTE (IBTE) is rather challenging due to the complexity involved in its integro-differential form \cite{ziman1960}. But linearized BTE solutions using momentum or self-energy relaxation time approximation (MRTA or SERTA) \cite{brunin2020phonon,giustino2017electron} can be used instead of IBTE for studying charge transport in semiconductors \cite{ponce2019hole,ponce2018towards,li2015electrical,brown2020band}. However, recent studies show some of these approximations may considerably underestimate carrier mobility in many semiconductors \cite{claes2022assessing,meng2019phonon,ma2018first}. While IBTE can produce results comparable to experimental data, its high computational demands make it impractical for routine applications.

In the ongoing quest for TE materials, silicides semiconductors can be particularly interesting due to their high TE conversion efficiency \cite{nozariasbmarz2017thermoelectric,borisenko2013semiconducting,vining_Silicides,liu2018eco,lovvik2020screening}. However, some materials in this class still need further treatment at a more accurate theoretical level for TE applications. For example, very recently, we predicted two silicides, Ca$_{2}$Si and Mg$_{2}$Si, as promising candidates for thin-film single-junction and multi-junction solar cells, respectively, based on advanced many-body theoretical calculations \cite{solet2024}. These silicides were classified as indirect and direct band gap semiconductors, exhibiting values of 0.6 and 0.96 eV, respectively, as determined using the mBJ method. In terms of TE properties, Mg$_{2}$Si has been studied extensively \cite{liu2012convergence,zaitsev2006highly,tani2005thermoelectric,yin2016optimization,johnson2018enhanced,noda1992temperature,zhang2008high,lovvik2020screening}, including many doping strategies with p-block elements \cite{tani2007thermoelectric_p,jung2011solid,kutorasinski2013calculating,tani2007thermoelectric_sb,bux2011mechanochemical,akasaka2008thermoelectric,yao2022lattice} at both experimental and computational levels \cite{nolas2007transport,muthiah2013conducting,tan2012multiscale,huan2016high,jung2010synthesis,kutorasinski2014importance,nozariasbmarz2016comparison}. Some of these studies have highlighted the high TE conversion efficiency achieved in $n$-type samples with specific dopant compositions \cite{liu2012convergence,zaitsev2006highly,tani2005thermoelectric,yin2016optimization,johnson2018enhanced,bux2011mechanochemical,kutorasinski2013calculating,yao2022lattice}. However, $p$-type samples have not shown great promise in delivering excellent performance \cite{akasaka2008thermoelectric,zhang2008high}. In comparison, relatively little attention has been given to Ca$_{2}$Si for evaluating its TE properties \cite{xiong2017structural,wen2018computational,lovvik2020screening,el2022theoretical}. Most of these TE studies on both silicides have relied solely on CRTA.

Based on this analysis, there is strong justification for selecting both silicides for further study. More accurate evaluations of their TE properties are needed beyond the CRTA method. Although some studies have explored the charge transport of Mg$_{2}$Si using approaches that go beyond CRTA, such as those incorporating EPI \cite{xia2022role,fan2018first}, these investigations are still pretty limited. For instance, Xia \textit{et al.} \cite{xia2022role} explored the impact of long-range e-p interactions and strain engineering on electrical conductivity, but their analysis was limited to 300 K and did not cover the complete TE properties. In another study, the TE properties of both silicides were explored using theory based on the deformation potential, accounting for acoustic phonon scattering as the primary scattering mechanism \cite{lovvik2020screening}. Similarly, Fan \textit{et al.} \cite{fan2018first} investigated the T-dependent (up to 900 K) TE behaviour of Mg$_{2}$Si and its solid solutions, but the properties they reported, such as $\sigma$, do not match meaningfully with experimental data. One possible reason for this mismatch could be that, in principle, accurate $\sigma$ calculations need careful testing to ensure enough wave vectors are sampled for both particles \cite{boussadoune2023electronic}, which was not conducted in their work. Another potential reason might be their use of an e-p interpolation procedure based on maximally localized Wannier functions (WFs). Unfortunately, generating WFs is not always straightforward and depends on a large number of parameters to obtain an appropriate set of WFs that effectively span the energy region of interest \cite{brunin2020phonon,agapito2018ab}. Therefore, one of the main aims of the current work is to make more accurate charge transport predictions using EPI as the main scattering mechanism at the first-principles level, crucial for gaining a better theoretical understanding of this compound for TE applications. Since experimental evidence for Ca$_{2}$Si is lacking, employing theoretical methods that have demonstrated accuracy in predicting results consistent with experiments for Mg$_{2}$Si offers a reliable way to validate the potential of Ca$_{2}$Si.     

Motivated by this interesting background, we employ a combination of first-principles calculations and BTE under different RTAs to study the transport properties in Ma$_{2}$Si and Ca$_{2}$Si. In Sec.~\ref{sec:methods}, we briefly review the methodology and fundamental theory underlying this work. Subsection~\ref{sec:2a} provides an overview of e-p SE and expresses the carrier relaxation time within SERTA and MRTA in this framework. The theory behind the calculation of quasiparticle (QP) corrections in electronic structure or EPR due to e-p coupling is discussed in Subsection~\ref{sec:2ba}. This subsection also focuses on various phonon-limited charge transport within electron Boltzmann theory. Phonon transport within phonon-phonon interaction (PPI) using the linearized phonon BTE is covered in Subsection~\ref{sec:2c}. Subsection~\ref{sec:2d} describes the computational methods employed in our first-principles calculations. Further, In Sec.~\ref{sec:result}, we present and discuss our results for Mg$_{2}$Si and Ca$_{2}$Si, focusing on phonon-induced renormalization of the electronic gap and the T-dependent electron mobility with a detailed convergence study at 300 K. This section also examines all TE transport coefficients, and then $z$T is discussed in various modeling conditions. Finally, Sec.~\ref{sec:concl} concludes the paper.  

\section{THEORY AND METHODS} \label{sec:methods}

\subsection{\label{sec:2a}Electron-phonon self-energy}
From the lowest order in perturbation theory \cite{giustino2017electron}, the e-p SE, $\Sigma^{\mathrm{ep}}_{n\textbf{k}}$, can be computed by the sum of the frequency dependent Fan-Migdal (FM) and the static and Hermitian Debye-Waller (DW) part,
\begin{eqnarray} \label{eq:se}
\Sigma^{\mathrm{ep}}_{n\textbf{k}}(\mathrm{\omega, T}) = \Sigma^{\mathrm{FM}}_{n\textbf{k}}(\mathrm{\omega, T}) +  \Sigma^{\mathrm{DW}}_{n\textbf{k}}(\mathrm{T}),                                 
\end{eqnarray} 
We briefly discuss each term. The \textit{ab initio} perturbative methods implement the diagonal elements of FM-SE matrix term in the Kohn-Sham (KS) basis set as \cite{brunin2020phonon}; 
\begin{eqnarray} \label{eq:fmse}
    \begin{gathered}
        \Sigma^{\mathrm{FM}}_{n\textbf{k}}(\mathrm{\omega, T}) = \sum_{m,\nu} \int_{\mathrm{BZ}}\frac{d\textbf{q}}{\Omega_{\mathrm{BZ}}} \Big|\mathrm{g}_{mn\nu}(\textbf{k},\textbf{q})\Big|^{2}\times 
         \\
         \Biggl[ \frac{n_{\textbf{q}\nu}+f_{m\textbf{k}+\textbf{q}}}{\omega - \varepsilon_{m\textbf{k}+\textbf{q}} + \omega_{\textbf{q}\nu} + i\eta}  + \frac{n_{\textbf{q}\nu} + 1 - f_{m\textbf{k}+\textbf{q}}}{\omega - \varepsilon_{m\textbf{k}+\textbf{q}} - \omega_{\textbf{q}\nu} + i\eta}\Biggl],            
    \end{gathered}
\end{eqnarray} 
where $n_{\textbf{q}\nu}(\mathrm{T})$ and $f_{m\textbf{k}+\textbf{q}}(\mathrm{T, \varepsilon_{F}})$ are, respectively, the Bose-Einstein and Fermi-Dirac occupation numbers at physical temperature T, and $\mathrm{\varepsilon_{F}}$ is the Fermi level. The integration is performed over the Brillouin zone (BZ) volume of $\Omega_{\mathrm{BZ}}$ for the phonon wave vectors and $\eta$ is a small positive real parameter. The e-p matrix elements $\mathrm{g}_{mn\nu}(\textbf{k},\textbf{q})$ represent the probability amplitude for scattering from an unperturbed Bloch state $|n\textbf{k}\rangle$, with band index $n$ and crystal momentum \textbf{k}, to $|m\textbf{k}+\textbf{q}\rangle$, due to the absorption or emission of a phonon with mode index $\nu$, wave-vector \textbf{q}, and energy $\omega_{\textbf{q}\nu}$ \cite{giustino2017electron},
\begin{eqnarray} \label{eq:gmn}
\mathrm{g}_{mn\nu}(\textbf{k},\textbf{q}) = \langle m\textbf{k}+\textbf{q}\big|\Delta_{\textbf{q}\nu}\mathrm{V^{KS}}\big|n\textbf{k}\rangle,                    
\end{eqnarray}
where $\Delta_{\textbf{q}\nu}$V$^{\mathrm{KS}}$ is the phonon-induced first-order variation in the self-consistent KS potential and defined as $e^{i\textbf{q}.\textbf{r}}\Delta_{\textbf{q}\nu}\upsilon^{\mathrm{KS}}(\textbf{r})$. The lattice periodic function $\Delta_{\textbf{q}\nu}\upsilon^{\mathrm{KS}}(\textbf{r})$ is expressed as,
\begin{eqnarray} \label{eq:latt}
\Delta_{\textbf{q}\nu}\upsilon^{\mathrm{KS}}(\textbf{r}) = \frac{1}{\sqrt{2\omega_{\textbf{q}\nu}}}\sum_{\kappa\alpha}\frac{e_{\kappa\alpha,\nu}(\textbf{q})}{\sqrt{\mathrm{M_{\kappa}}}}\partial_{\kappa\alpha,\textbf{q}}\upsilon^{\mathrm{KS}}(\textbf{r}),            
\end{eqnarray}
where $e_{\kappa\alpha,\nu}(\textbf{q})$ is phonon eigenvector for $\alpha$th Cartesian component of $\kappa$th ion with mass M$_{\kappa}$ in a unit cell. The $\partial_{\kappa\alpha,\textbf{q}}\upsilon^{\mathrm{KS}}(\textbf{r}$) is first order derivative of KS potential experienced by the electrons at a given ($\kappa\alpha$, $\textbf{q}$) perturbation, which is obtained from density functional perturbation theory (DFPT) by solving the self-consistently of Sternheimer equations \cite{baroni2001phonons}. 
  
The frequency-independent DW term in Eq.~\hyperref[eq:se]{\eqref{eq:se}} 
\begin{eqnarray}
\Sigma^{\mathrm{DW}}_{n\textbf{k}}(\mathrm{T}) = \sum_{\textbf{q}\nu m}\Big(2n_{\textbf{q}\nu}(\mathrm{T})+1\Big)\frac{\mathrm{g}_{mn\nu}^{2,\mathrm{DW}}(\textbf{k},\textbf{q})}{\varepsilon_{n\textbf{k}}-\varepsilon_{m\textbf{k}}},                              
\end{eqnarray}
depends on the second order derivative of the KS potential with respect to the ion displacements, represented by the effective matrix elements g$_{mn\nu}^{2,\mathrm{DW}}(\textbf{k},\textbf{q})$, which are generally challenging to compute. To circumvent this difficulty, the equation above is evaluated within the rigid-ion approximation in terms of the first-order g$_{mn\nu}(\textbf{k},\textbf{q})$ matrix elements \cite{giustino2017electron,ponce2014temperature,allen1976theory}.

Next, the different approximations exist to obtain the electron scattering lifetime $\uptau_{n\textbf{k}}$ due to EPI. One can obtain lifetime from the imaginary part of the FM-SE [Eq.~\hyperref[eq:fmse]{\eqref{eq:fmse}}] at the KS energy $\varepsilon_{n\textbf{k}}$ \cite{giustino2017electron,brunin2020phonon}. Within SERTA, one defines in $\eta$ $\to$ $0^{+}$ limit,
\begin{eqnarray} \label{eq:serta}
    \begin{gathered}
        \frac{1}{\uptau_{n\textbf{k}}} =2\lim_{\eta \to 0^{+}} \mathrm{Im}\big[\Sigma^{\mathrm{FM}}_{n\textbf{k}}(\varepsilon_{n\textbf{k}})\big] = 2\pi \sum_{m,\nu} \int_{\mathrm{BZ}} \frac{d\textbf{q}}{\Omega_{\mathrm{BZ}}} \Big|\mathrm{g}_{mn\nu}(\textbf{k},\textbf{q})\Big|^{2} 
        \\
        \times\Bigl[ (n_{\textbf{q}\nu} + f_{m\textbf{k}+\textbf{q}}) \delta(\varepsilon_{n\textbf{k}} - \varepsilon_{m\textbf{k}+\textbf{q}} + \omega_{\textbf{q}\nu}) 
        \\ 
        + (n_{\textbf{q}\nu} + 1 -f_{m\textbf{k}+\textbf{q}})\delta(\varepsilon_{n\textbf{k}} - \varepsilon_{m\textbf{k}+\textbf{q}} - \omega_{\textbf{q}\nu})\Bigl].        
    \end{gathered}
\end{eqnarray} 
The first (second) energy conserving $\delta$-function in the equation represents the phonon absorption (emission) process. Another widely used and more accurate approximation is the MRTA \cite{giustino2017electron,ziman1960}, which accounts for the relative change in electron velocity or momentum loss in the scattering processes. Thus, $\uptau_{n\textbf{k}}$ is calculated by multiplying a velocity factor $(1 - \hat{\textbf{v}}_{n\textbf{k}}.\hat{\textbf{v}}_{m\textbf{k}+\textbf{q}})$ in Eq.~\hyperref[eq:serta]{\eqref{eq:serta}}. 

\subsection{\label{sec:2ba}EPR and transport parameters}
The QP energy due to e-p coupling can be estimated by real part of SE at the bare band energy $\varepsilon_{n\textbf{k}}$ using Eq.~\hyperref[eq:se]{\eqref{eq:se}} through two different approaches. In the first approach, the EPR energy at T using on-the-mass-shell (OTMS) approximation is given by \cite{cannuccia2012zero}, 
\begin{eqnarray}
\varepsilon_{n\textbf{k}}(\mathrm{T}) =  \varepsilon_{n\textbf{k}} + \mathrm{Re}\Big[\Sigma_{n\textbf{k}}^{\mathrm{ep}}(\varepsilon_{n\textbf{k}},\mathrm{T})\Big].           
\end{eqnarray}
However from the solution of linearized-QP-equation (LQE), the above equation is changed via renormalization factor Z$_{n\textbf{k}}$,
\begin{eqnarray}
\varepsilon_{n\textbf{k}}(\mathrm{T}) =  \varepsilon_{n\textbf{k}} + \mathrm{Z}_{n\textbf{k}} \mathrm{Re}\Big[\Sigma_{n\textbf{k}}^{\mathrm{ep}}(\varepsilon_{n\textbf{k}},\mathrm{T})\Big],            
\end{eqnarray}
and,
\begin{eqnarray} \label{eq:znk}
\mathrm{Z}_{n\textbf{k}} = \Big(1 - \mathrm{Re}\Big[\frac{\partial\Sigma_{n\textbf{k}}^{\mathrm{ep}}}{\partial\varepsilon}\Big]\Big|_{\varepsilon = \varepsilon_{n\textbf{k}}} \Big)^{-1}.        
\end{eqnarray}

The ZPR of the excitation energy is calculated as the difference between the energy obtained from the equations above at T = 0 and the bare KS eigenvalue.

For electrical transport properties, the electron Boltzmann theory is used. An alternative formulation of the linearized BTE in the RTA employs the transport distribution function integrating over \textbf{k}-space for bands $n$ \cite{brunin2020phonon,ashcroft1976},
\begin{eqnarray} \label{eq:trans}
 \mathcal{L}_{\alpha\beta}^{(m)} = -\sum_{n}\int\frac{d \bf k}{\Omega_{\mathrm{BZ}}}\mathrm{v}_{n\textbf{k},\alpha}\mathrm{v}_{n\textbf{k},\beta} \uptau_{n\textbf{k}}(\varepsilon_{n\textbf{k}}-\varepsilon_{\mathrm{F}})^{m}\frac{\partial f}{\partial \varepsilon }\Bigg|_{\varepsilon_{n\textbf{k}}} \quad
\end{eqnarray}
where v$_{n\textbf{k},\alpha}$ is the $\alpha$th Cartesian component of the matrix element for the electron velocity operator \textbf{v}$_{n\textbf{k}}$ in state $|n\textbf{k}\rangle$. Within this definition, the Cartesian components of transport coefficients related to TE have familiar forms as \cite{madsen2018boltztrap2};    
\begin{eqnarray} \label{eq:coeff}
    \begin{gathered}
        \sigma_{\alpha\beta} = \mathcal{L}^{(0)}_{\alpha\beta}, \quad \mathrm{S}_{\alpha\beta} = \frac{1}{\mathrm{qT}}\frac{\mathcal{L}^{(1)}_{\alpha\beta}}{\mathcal{L}^{(0)}_{\alpha\beta}},
        \\
        (\kappa_{\mathrm{e}})_{\alpha\beta} = \frac{1}{\mathrm{q^{2}T}}\bigg(\frac{(\mathcal{L}^{(1)}_{\alpha\beta})^{2}}{\mathcal{L}^{(0)}_{\alpha\beta}} - \mathcal{L}^{(2)}_{\alpha\beta}\bigg).         
    \end{gathered}
\end{eqnarray}

The $\sigma$ can be expressed in terms of the contributions from electrons and holes as: $\sigma$ = $n_{\mathrm{e}}\mu_{\mathrm{e}}$ + $n_{\mathrm{h}}\mu_{\mathrm{h}}$ \cite{brunin2020phonon}. In this expression; n$_{\mathrm{e}}$, $\mu_{\mathrm{e}}$ and $n_{\mathrm{h}}$, $\mu_{\mathrm{h}}$ refer to the particle density and mobility of electrons and holes, respectively, and can be obtained as,
\begin{eqnarray} \label{eq:ne}
n_{\mathrm{e}} = \sum_{n\in \mathrm{CB}}\int\frac{d\textbf{k}}{\Omega_{\mathrm{BZ}}}f_{n\textbf{k}}, \quad  \mu_{\mathrm{e}} = \frac{\mathcal{L}^{(0)}_{n\in \mathrm{CB}}}{n_{\mathrm{e}}\Omega}
\end{eqnarray}
where the summation is restricted to states in the conduction bands ($n$$\in$CB). $\Omega$ is the volume of the crystalline unit cell. Similarly, hole density and mobility can be calculated in an analogous manner. 

\subsection{\label{sec:2c}Lattice thermal conductivity}
In this paper, $\kappa_{\mathrm{ph}}$ is computed from the imaginary part of phonon SE, $\Gamma_{\lambda}(\omega_{\lambda})$ ($\lambda$ = $\textbf{q}\nu$), by considering only PPI \cite{togo2015distributions}. The quantity $\Gamma_{\lambda}(\omega_{\lambda})$ is calculated with anharmonic third-order force constant from the Fermi’s golden rule as \cite{togo2015distributions};
\begin{eqnarray} \label{eq:ppi}
\Gamma_{\lambda}(\omega) = \frac{18\pi}{\hbar^{2}}\sum_{\lambda^{'}\lambda^{''}} \Big| \Phi_{-\lambda\lambda^{'}\lambda^{''}}\Big|^{2}\biggl\{\bigl(n_{\lambda^{'}}+ n_{\lambda^{''}}+ 1\bigl) \nonumber \\ \times \delta\bigl(\omega - \omega_{\lambda^{'}} - \omega_{\lambda^{''}}\bigl) + \bigl(n_{\lambda^{'}} - n_{\lambda^{''}}\bigl) \quad\nonumber \\\times \Bigl[\delta\bigl(\omega + \omega_{\lambda^{'}} - \omega_{\lambda^{''}}\bigl) - \delta\bigl(\omega - \omega_{\lambda^{'}} + \omega_{\lambda^{''}}\bigl)\Bigl] \biggl\}, \quad
\end{eqnarray}
with $n_{\lambda}$ being Bose–Einstein occupation number at thermal equilibrium. The term $\Phi_{-\lambda\lambda^{'}\lambda^{''}}$ is the PPI strength, which is estimated from second-order perturbation theory for the three phonon process involving modes $\lambda$, $\lambda^{'}$ and $\lambda^{''}$. Twice the value of Eq.~\hyperref[eq:ppi]{\eqref{eq:ppi}} is known as the phonon linewidth for $\lambda$, and its reciprocal estimates the phonon lifetime $\uptau_{\lambda}$ \cite{togo2015distributions}. In fact $\kappa_{\mathrm{ph}}$ and $\uptau_{\lambda}$ are straightforwardly related through the single-mode relaxation-time approximation (SMRTA) \cite{srivastava2019}. Therefore, the $\kappa_{\mathrm{ph}}$ tensor is obtained by solving linearized phonon BTE under SMRTA method as \cite{togo2015distributions};
\begin{eqnarray} \label{eq:kappaph}
\kappa_{\mathrm{ph}} = \frac{1}{\mathrm{NV_{0}}}\sum_{\lambda}\mathrm{C}_{\lambda}\textbf{v}_{\lambda}\otimes\textbf{v}_{\lambda}\uptau^{\mathrm{SMRTA}}_{\lambda}.
\end{eqnarray}
where V$_{0}$ and N are the volume of the unit cell and number of unit cells in the system, respectively. $\textbf{v}_{\lambda}$ and C$_{\lambda}$ represent the mode-dependent phonon group velocity and specific heat, respectively, and can be obtained directly from the solution of eigenvalue problem \cite{srivastava2019,solet2023ab}. It is assumed that $\uptau^{\mathrm{SMRTA}}_{\lambda}$ = $\uptau_{\lambda}$ for the computation of $\kappa_{\mathrm{ph}}$ in the above equation.

\subsection{\label{sec:2d}Computational details}
All calculations from first-principles are performed using an optimized norm-conserving Vanderbilt pseudopotential \cite{hamann2013optimized} with the Perdew-Burke-Ernzerhof (PBE) \cite{perdew1996generalized} exchange-correlation functional in the Abinit software \cite{gonze2002first,gonze2020abinit}. The ground state electronic energies are obtained from DFT, while phonon frequencies and related potentials derivatives are derived from DFPT \cite{gonze1997dynamical,baroni2001phonons} with a plane wave cutoff kinetic energy of 40 Hartree for both compounds. The ground state electronic density is calculated on a converged $\Gamma$-centered irreducible \textbf{k}-mesh (IBZ$_{\textbf{k}}$) of 16 $\times$ 16 $\times$ 16. A $\Gamma$-centered irreducible coarse \textbf{q}-mesh (IBZ$_{\textbf{q}}$) of 8 $\times$ 8 $\times$ 8 is employed to achieve an accurate Fourier interpolation of the e-p scattering potentials in Eq.~\hyperref[eq:latt]{\eqref{eq:latt}}. Then the DFPT scattering potentials for ZPR and temperature-dependent band-gap calculations is performed on a well converged mesh of 48 $\times$ 48 $\times$ 48 \textbf{q} and \textbf{k}-points. A small complex shift value of 0.01 eV is used to avoid divergences in the denominator of the FM SE. The results are obtained using the Sternheimer approach on 35 well-converged bands for both compounds \cite{gonze2011theoretical}.

In calculating mobility, a dense sampling of phonon (\textbf{q}-point) and electron (\textbf{k}-point) wave vectors in the BZ is required. A large number of Bloch states are computed non-self consistently from an initial set of IBZ$_{\textbf{k}}$-points, while a finer \textbf{q} mesh is obtained by Fourier interpolating the scattering potentials onto a coarse IBZ$_{\textbf{q}}$-mesh. To reduce computational burden, we have employed a double-grid (DG) technique, in which density of \textbf{k} and \textbf{q} points is obtained on the same mesh, and the \textbf{q}-point density is calculated on a grid with double the size to describe the phonon absorption/emission terms in Eq.~\hyperref[eq:serta]{\eqref{eq:serta}}. The states $|n\textbf{k}\rangle$ within the energy window from conduction band minima (CBM) to 0.27 (0.3) eV in Ca$_{2}$Si (Mg$_{2}$Si) are used in Eq.~\hyperref[eq:se]{\eqref{eq:se}} to compute the electron lifetime. For the iterative process in the IBTE, the convergence criterion is set to be 10$^{-5}$. To obtain intrinsic $\mu_{\mathrm{e}}$ from Eq.~\hyperref[eq:ne]{\eqref{eq:ne}}, we introduce a small electron doping of $10^{15}$~{\cm} in the conduction band for both materials. 

The TE transport properties at CRTA [using $\uptau_{n\textbf{k}}$ = $\uptau$ = constant in Eq.~\hyperref[eq:trans]{\eqref{eq:trans}}] have been calculated using the BoltzTraP code \cite{madsen2006}. Furthermore, $\kappa_{\mathrm{ph}}$ is calculated using PPI under SMRTA based on phonon Boltzmann theory, as implemented in the Phono3py code \cite{togo2015distributions}. Harmonic and anharmonic inter-atomic force constants are determined within DFT using Abinit. Both force constants are computed within the unit cells of 2 $\times$ 2 $\times$ 2 supercell using the finite atomic displacement method, considering interactions up to the third-nearest neighbors for anharmonic force constants to keep the number of supercell calculations manageable. A cutoff energy for plane waves of 25 Hartree, a 4 $\times$ 4 $\times$ 4 \textbf{k}-mesh grid, and a force convergence criteria of 5 $\times$ $10^{-8}$ Hartree/Bohr are selected for the force calculations in the supercell. Finally for $\kappa_{\mathrm{ph}}$ calculations, a phonon wave vectors grid of 23 $\times$ 23 $\times$ 23 is used to integrate over the BZ in \textbf{q} space.

\section{Results and discussion} \label{sec:result}
\subsection{\label{sec:3a}Electronic band-gap renormalization}

\begin{figure}
\includegraphics[width=7.8cm, height=10.0cm]{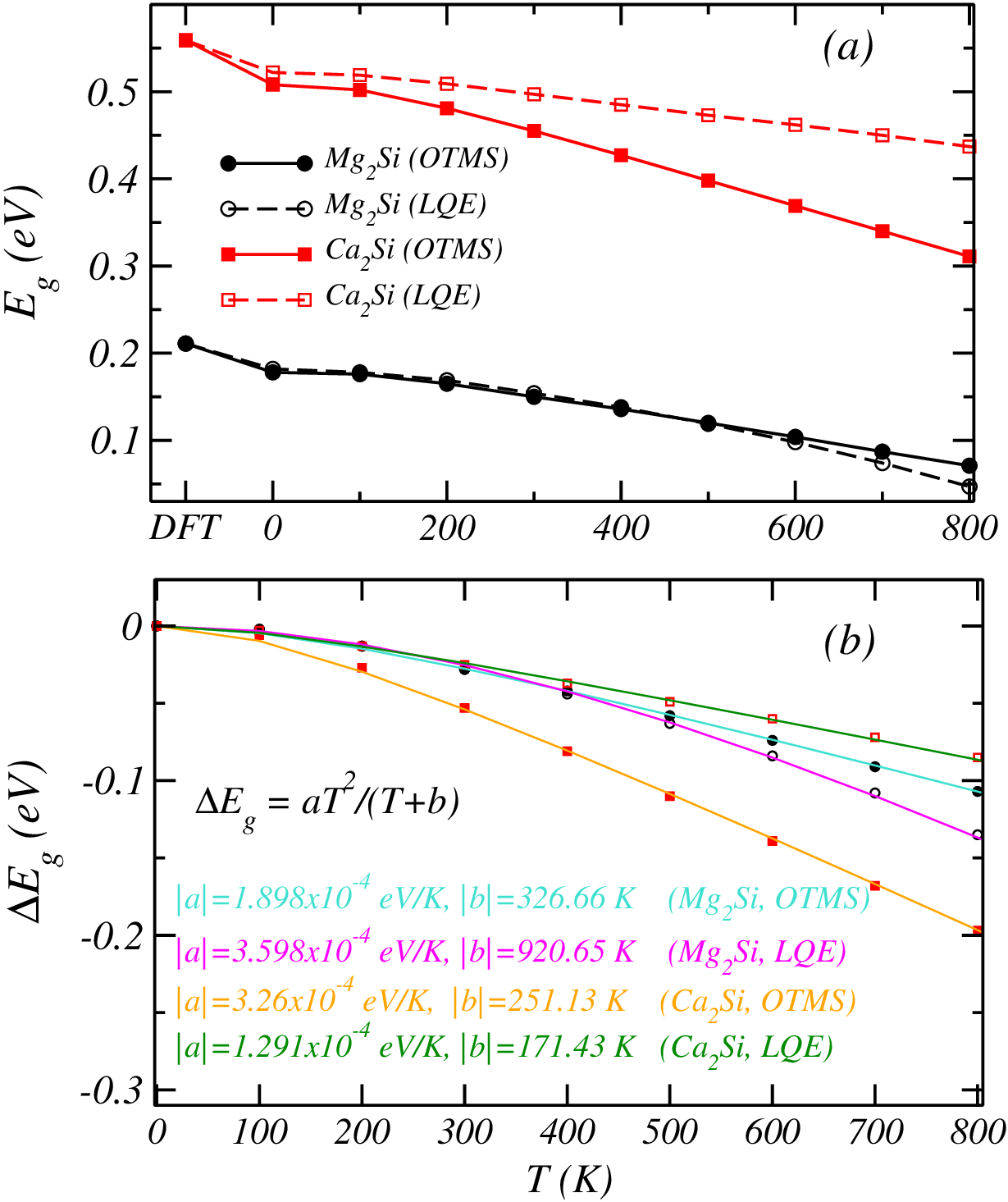} 
\caption{ (a) DFT, ZPR correction in band gap at 0 K and temperature dependence of indirect (for Mg$_{2}$Si) and direct (for Ca$_{2}$Si) band-gaps from on-the-mass-shell (OTMS) [solid line] and linearized-QP-equation (LQE) [dashed line] methods. (b) The absolute values of coefficients $|a|$ and $|b|$ are obtained from non-linear fittings of Varshni equation of $\Delta$E$_{\mathrm{g}}$ = $a$T$^{2}$/(T+$b$).}
\label{fig:gap}
\end{figure} 

\begin{table}
\caption{Band gaps (eV) from on-the-mass-shell (OTMS) and linearized-QP-equation (LQE) solutions for Mg$_{2}$Si and Ca$_{2}$Si at different temperatures together with zero-point renormalization (ZPR) and DFT (without EPR).}
\resizebox{0.47\textwidth}{!}{
\begin{tabular}{@{\extracolsep{\fill}} c c c c c c c c c c}
\hline\hline 
Temperature&& \multicolumn{2}{c}{Mg$_{2}$Si (indirect gap)} &  & \multicolumn{2}{c}{Ca$_{2}$Si (direct gap)} &  & \\
\cline{2-4} \cline{6-9} 
(in K)  &&OTMS &LQE  &&OTMS &LQE && \\
\hline   
ZPR       & & -0.033   & -0.029   & & -0.051   & -0.037   &&  \\
Without EPR & &  0.211   &  0.211   & &  0.559   &  0.559   &&  \\
0           & &  0.178   &  0.182   & &  0.508   &  0.522   &&  \\
100         & &  0.176   &  0.178   & &  0.502   &  0.519   &&  \\
200         & &  0.165   &  0.169   & &  0.481   &  0.509   &&  \\
300         & &  0.150   &  0.154   & &  0.455   &  0.497   &&  \\
400         & &  0.136   &  0.138   & &  0.427   &  0.485   &&  \\
500         & &  0.120   &  0.119   & &  0.398   &  0.473   &&  \\
600         & &  0.104   &  0.098   & &  0.369   &  0.462   &&  \\
700         & &  0.087   &  0.074   & &  0.340   &  0.450   &&  \\
800         & &  0.071   &  0.047   & &  0.311   &  0.437   &&  \\

\hline \hline 
\end{tabular}}
\label{table:bandgap}
\end{table} 

The effect of temperature on the electronic structure is crucial for understanding materials used in TE generators. Figure~\hyperref[fig:gap]{\ref{fig:gap}} illustrates the fundamental band gap calculated using PBE for both silicides, presenting values without the influence of EPR$-$referred to as DFT$-$and with EPR effects included. The latter accounts for both ZPR and T dependence scenarios, using OTMS and LQE approaches. While DFT typically treats nuclei as classical particles clamped at their equilibrium positions, it is essential to recognize that even at absolute zero, atoms in solids exhibit quantum zero-point motion. According to Allen, Heine, and Cardona (AHC) theory \cite{cardona2006superconductivity}, interactions between electrons and phononic fields cause a renormalization of electron energy levels even at zero Kelvin. This adjustment in ground-state energy is known as the ZPR, defined as E$_{\mathrm{g}}$(T = 0)-E$_{\mathrm{g}}$(static). 

Figure~\hyperref[fig:gap]{\ref{fig:gap}(a)} presents the ZPR of the KS band gap, which is estimated to be $\sim$-33 (-29) meV and -51 (-37) meV for Mg$_{2}$Si and Ca$_{2}$Si, respectively, using the OTMS (LQE) method. This correction reduces the DFT-PBE gap from 0.211 to 0.178 (0.182) for Mg$_{2}$Si and from 0.559 to 0.508 (0.522) for Ca$_{2}$Si. As shown in Table~\ref{table:bandgap}, the band gap decreases with increasing T from both methods and for both silicides, which can be explained by Varshni’s effect \cite{giustino2017electron,varshni1967temperature}. At 300 K, the calculated band gap values are $\sim$0.15-0.154 eV for Mg$_{2}$Si and $\sim$0.455-0.497 eV for Ca$_{2}$Si. Furthermore, the band gap values decrease by almost 66-78\% for Mg$_{2}$Si and 22-24\% for Ca$_{2}$Si at 800 K compared to the DFT gap values. This indicates that self-energy renormalization leads to a larger change in the band gap for Mg$_{2}$Si than Ca$_{2}$Si, which can affect electron transport phenomena in these materials. A previous study on ZPR correction for the mBJ band gap within the EPR framework, accounting for both lattice expansion and lattice vibrations in Mg$_{2}$Si, reported a value of almost -40 meV \cite{yao2023temperature}. This suggests that lattice vibrations are the dominate factor in ZPR correction for Mg$_{2}$Si. Beyond ZPR, temperature-dependent band gap renormalization is also dominated by the EPI, as observed for Mg$_{2}$Si in Ref. \cite{yao2023temperature} and other semiconductors such as rutile TiO$_{2}$ \cite{wu2018first}. The Z-factor [in Eq.~\hyperref[eq:znk]{\eqref{eq:znk}}] has little impact on the band gap of Mg$_{2}$Si, while its influence becomes more pronounced with increasing T in Ca$_{2}$Si, compared to the gap obtained by OTMS, can be noticed in Fig.~\hyperref[fig:gap]{\ref{fig:gap}(a)}. This indicates that change in Re$\Sigma^{\mathrm{ep}}_{n\textbf{k}}$ with respect to KS energy is more significant in Ca$_{2}$Si than in Mg$_{2}$Si. Furthermore, at higher temperatures, including the Z-factor decreases the gap value relative to OTMS in Mg$_{2}$Si, whereas it enhances the gap value in Ca$_{2}$Si, resulting in opposite impacts of the Z-factor on the two silicides.  

To gain a deeper understanding of how band gaps vary with T in both OTMS and LQE methods, the Varshni equation of $\Delta$E$_{\mathrm{g}}$(T) = $a$T$^{2}$/(T+$b$) is analyzed. Here, $\Delta$E$_{\mathrm{g}}$(T) represents the change in the band gap at T relative to its value at absolute zero (0 K), i.e., E$_{\mathrm{g}}$(T)-E$_{\mathrm{g}}$(T = 0). This change is plotted for both compounds in Fig.~\hyperref[fig:gap]{\ref{fig:gap}(b)}. Meanwhile, the fitting coefficients with absolute values ($|a|$ and $|b|$) are determined by applying the equation, with the band gap at 0 K set as the reference point. Specifically, a larger $|a|$ indicates a stronger change in $\Delta$E$_{\mathrm{g}}$(T) with T, while a smaller value of $|b|$ represents a more linear relationship between $\Delta$E$_{\mathrm{g}}$(T) and T. For Mg$_{2}$Si, the absolute values of $|a|$ are $\sim$1.898 and 3.598 $\times$ $10^{-4}$ eV/K using OTMS and LQE methods, respectively. The estimated $|b|$ values using respective methods are $\sim$326.66 K and 920.65 K. This means that within the studied T range, including Z-factor significantly increases both coefficients, leading to a comparatively larger variation in band gap with T, as compared to OTMS. In contrast, for Ca$_{2}$Si, both $|a|$ and $|b|$ decrease from $\sim$3.26 $\times$ $10^{-4}$ eV/K to $\sim$1.291 $\times$ $10^{-4}$ eV/K and from $\sim$251.133 K to 171.435 K, respectively, when the Z-factor is included over OTMS. This suggests that for Ca$_{2}$Si, the inclusion of Z-factor reduces the sensitivity of the band gap variation to T, leading to a more linear dependence of $\Delta$E$_{\mathrm{g}}$(T) on T. This behavior of the Z-factor is completely opposite to what has been observed for Mg$_{2}$Si. This investigation into the T-dependent band gap highlights the critical role of the renormalization factor Z$_{n\textbf{k}}$ for both semiconducting silicides. The information of the Z-factor is also directly related to the observation of well-defined QP excitations in materials, which is crucial for gaining a deeper understanding of EPR physics in silicides.

\begin{figure*}[ht]
\includegraphics[width=15.5cm, height=6.5cm]{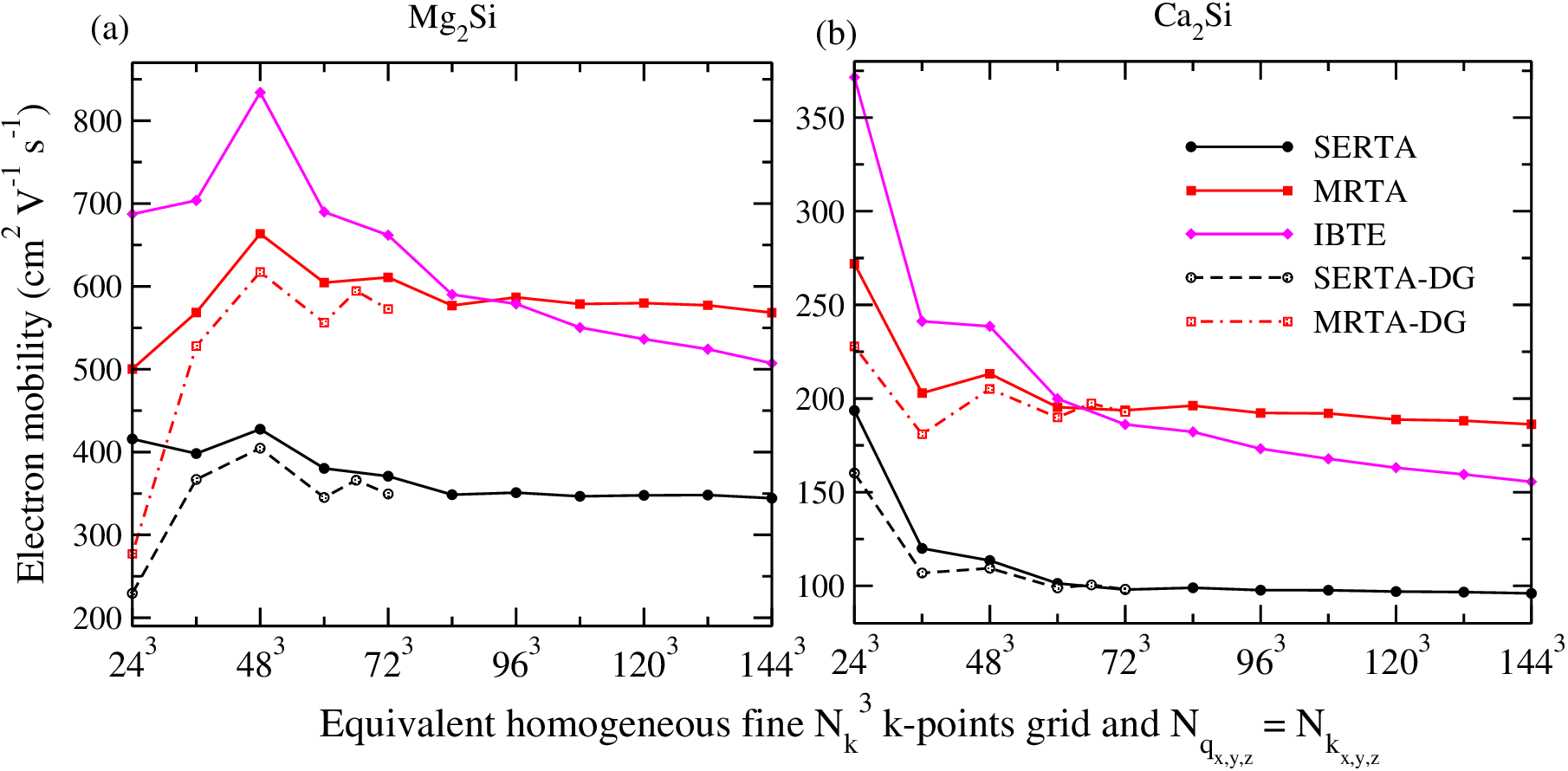} 
\caption{Convergence of the electron mobility at 300 K with respect to the Brillouin zone sampling of electron (\textbf{k}) and phonon (\textbf{q}) wave vectors, using methods based on SERTA and MRTA [with and without using double grid (DG)], and the iterative solution of the BTE (IBTE).} 
\label{fig:mob_con}
\end{figure*}

\subsection{\label{sec:3b}Convergence test for mobility}
Charge carrier mobility is one of the crucial property for semiconducting technologies such as TE, solar cells and light-emitting devices. One of the main challenges in accurately calculating the mobility is the requirement for a very dense sampling of phonon (\textbf{q}-point) and electron (\textbf{k}-point) wave vectors in the BZ \cite{brunin2020phonon}. To address this, we first conduct a convergence study of electron mobility $\mu_{\mathrm{e}}$ at 300 K by varying the fine \textbf{q-} and \textbf{k-}points grid. Additionally, we assess the impact of the DG method on the convergence rate. Figure~\ref{fig:mob_con} presents the dependence of $\mu_{\mathrm{e}}$ on the equivalent homogeneous dense \textbf{q} (N$_{\mathrm{q_{x,y,z}}}$) and \textbf{k} (N$_{\mathrm{k_{x,y,z}}}$) points grid used in calculating $\uptau_{n\textbf{k}}$ in Eq.~\hyperref[eq:serta]{\eqref{eq:serta}} and subsequently in the integration of Eq.~\hyperref[eq:trans]{\eqref{eq:trans}}. 

Figure~\hyperref[fig:mob_con]{\ref{fig:mob_con}(a)} compares the convergence study of $\mu_{\mathrm{e}}$ for Mg$_{2}$Si obtained from the SERTA and MRTA methods, both with and without DG, as well as the IBTE method. All convergence curves exhibit a similar trend: at lower \textbf{k}-point densities, the mobility increases with rising \textbf{k}-point density up to $48\times 48\times 48$. This happens because the sampling of \textbf{k}-points in the conduction band becomes denser near the CBM, where the integration in Eq.~\hyperref[eq:trans]{\eqref{eq:trans}} reaches its maximum. However, beyond a large enough \textbf{k}-point density of $48\times 48\times 48$, the mobility begins to decrease due to the increased scattering channels that arise from the higher \textbf{q}-point densities. The SERTA (MRTA) $\mu_{\mathrm{e}}$, calculated at $84\times 84\times 84$ \textbf{k} and \textbf{q} grid is $\sim$350 (577) {\cvs}, with convergence less than 2\% achieved for a grid size up to $144\times 144\times 144$. In contrast, the $\mu_{\mathrm{e}}$ computed from IBTE for the same mesh of electrons and phonons exhibits a very slow converging rate. Even at large meshes, from $132^{3}$ to $144^{3}$, full convergence is still not achieved, with an error of almost 5\%. So based on the convergence curve trend, we can say that IBTE solution requires a very high density of wave-vectors for both particles to achieve full convergence. In other words, it demands large computational resources, which limits its practicality for real-world applications. The $\mu_{\mathrm{e}}$ values calculated using the DG method for SERTA (SERTA-DG) and MRTA (MRTA-DG) are nearly identical to those obtained from the respective SERTA and MRTA methods. However, a significant advantage of the DG technique is its reduced computational time. Our converged $\mu_{\mathrm{e}}$ from SERTA using the DG method of $\sim$351 {\cvs} is well matched with the experimentally measured $\mu_{\mathrm{e}}$ of $\sim$350 {\cvs} for single crystal Mg$_{2}$Si at 300 K \cite{morris1958semiconducting}, while the MRTA and IBTE significantly overestimate the experimental value.

Figure~\hyperref[fig:mob_con]{\ref{fig:mob_con}(b)} illustrates the convergence test of $\mu_{\mathrm{e}}$ for Ca$_{2}$Si. First of all, we observe an overall decreasing trend in $\mu_{\mathrm{e}}$ as the \textbf{k} and \textbf{q} point densities increase from low (48$^{3}$) to dense (144$^{3}$) across all the studied methods. The well-converged value is achieved at grid of $60\times 60\times 60$ \textbf{k} and \textbf{q} points for both SERTA and MRTA, without and with applying the DG method. Oscillations are still observed in the MRTA results for \textbf{k} and \textbf{q} meshes denser than $60^{3}$, but these do not exceed 2\%. This indicates that a relatively small {\textbf{k}} and \textbf{q}-points grid is sufficient for Ca$_{2}$Si to obtain well-converged $\mu_{\mathrm{e}}$, especially compared to the large grids required for Mg$_{2}$Si. This can be understood by the band dispersion: the smaller effective mass (and thus more dispersive band) in Mg$_{2}$Si compared to Ca$_{2}$Si requires very high resolution in \textbf{k} space to accurately sample the small electron pocket around the CBM \cite{solet2024}. Once the \textbf{q} points grid becomes sufficiently dense to ensure convergence ($60\times 60\times 60$ mesh), the $\mu_{\mathrm{e}}$ obtained from the DG technique closely resembles values calculated without the DG method in both SERTA and MRTA. The values at this mesh are $\sim$100 {\cvs} using SERTA and $\sim$197 {\cvs} using MRTA. We have also noted as in Mg$_{2}$Si that the IBTE method converges more slowly than the other methods, reaching convergence within 3\% at a value of $\sim$163 {\cvs} with a grid of $144\times 144\times 144$ for both \textbf{k} and \textbf{q} points. Finally, for both materials, relatively low \textbf{k} and \textbf{q} points are generally sufficient to obtain a fully converged $\mu_{\mathrm{e}}$ within linearized BTE solution as opposed to the exact one. However, achieving convergence for exact solution within 2-3\% requires very large grids ($144\times 144\times 144$) for both electron and phonon wave vectors. 

\begin{figure}[ht]
\includegraphics[width=7.9cm, height=10.0cm]{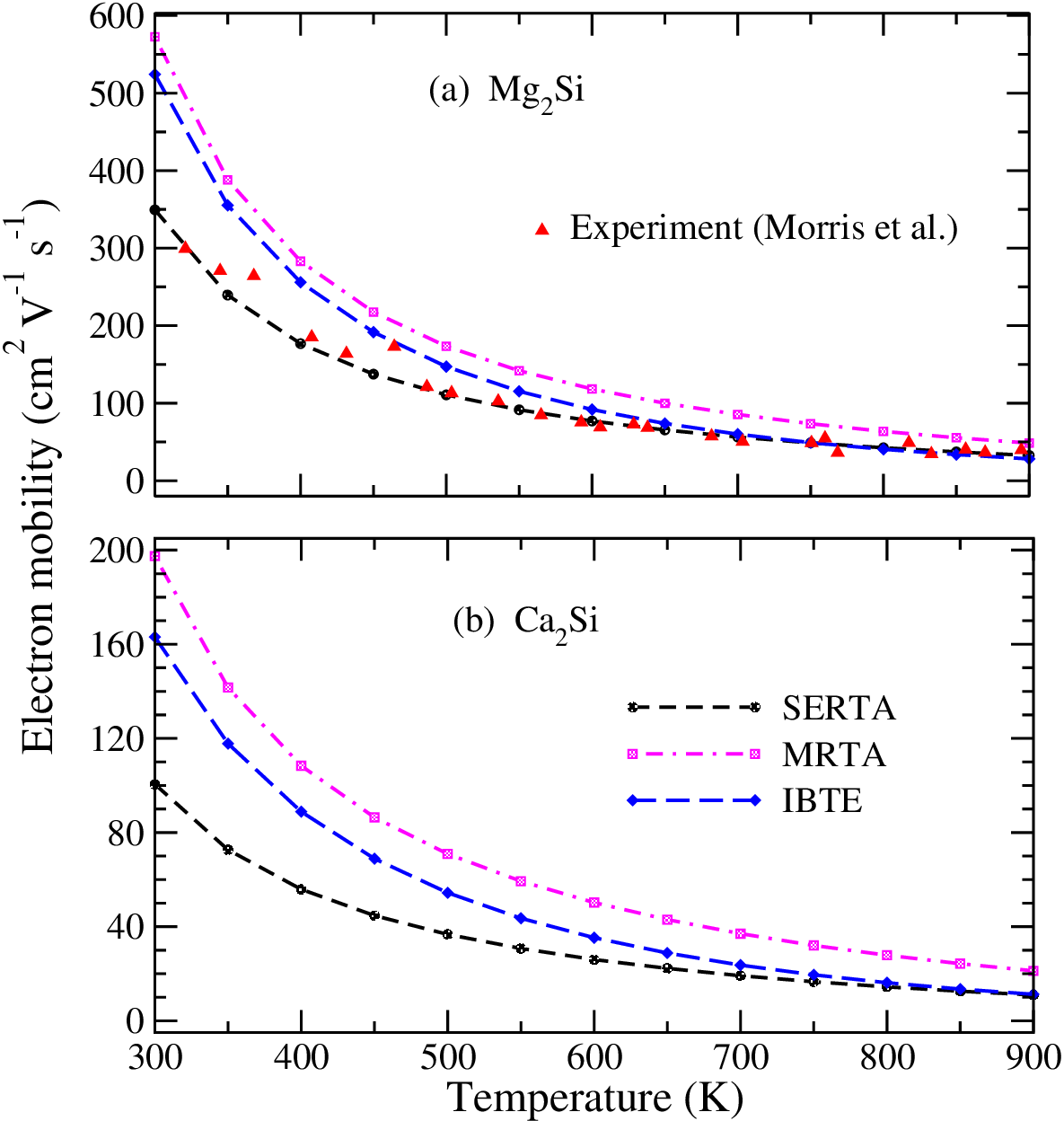} 
\caption{Temperature-dependent electron mobility of (a) Mg$_{2}$Si and (b) Ca$_{2}$Si, calculated using the linearized (SERTA, MRTA) and iterative BTE (IBTE). Experimental data for Mg$_{2}$Si by Morris \textit{et al.} \cite{morris1958semiconducting}, represented by triangles in panel (a), are included for comparison.} 
\label{fig:mob}
\end{figure} 

\subsection{\label{sec:3c}Results for the phonon-limited mobility}
Following a detailed convergence study on $\mu_{\mathrm{e}}$, Fig.~\ref{fig:mob} presents the well-converged $\mu_{\mathrm{e}}$ results for both silicides across the T range of 300-900 K. In both materials, $\mu_{\mathrm{e}}$ decreases with increasing T. The underlying physical reason for this behaviour is due to the availability of more states for scattering with electrons at higher temperatures. Figure~\hyperref[fig:mob]{\ref{fig:mob}(a)} compares the $\mu_{\mathrm{e}}$ results of Mg$_{2}$Si obtained from the SERTA, MRTA, and IBTE methods with experimental data for single-crystal Mg$_{2}$Si reported by Morris \textit{et al.} \cite{morris1958semiconducting}. In this figure, the SERTA method provides the best match for $\mu_{\mathrm{e}}$ compared to the experimental data. However, in high T region ($>$600 K) there is no significant difference between the $\mu_{\mathrm{e}}$ values calculated from SERTA and IBTE methods, matching well with experimental data. Throughout the T range studied, $\mu_{\mathrm{e}}$ from MRTA is consistently exceeded those from IBTE and the experimental values. However, as the T rises, this overestimation compared to the experiment (IBTE) goes down (up), and leads to a improved agreement with experimental data at higher temperatures. The calculated room-temperature values are $\sim$351, 573, and 524 {\cvs} from SERTA, MRTA, and IBTE, respectively, with SERTA showing very close agreement to the experimental value of $\sim$350 {\cvs} \cite{morris1958semiconducting}. At 900 K, these corresponding values drop to $\sim$33, 48, and 28 {\cvs}, which are well consistent with the experimental value of almost 38 {\cvs} \cite{morris1958semiconducting}. The same observation is also revealed in other semiconductors, such as silicon and GaAs, where SERTA predicts $\mu_{\mathrm{e}}$ close to the experimental results \cite{claes2022assessing,ma2018first}. 

Figure~\hyperref[fig:mob]{\ref{fig:mob}(b)} shows the calculated $\mu_{\mathrm{e}}$ values for Ca$_{2}$Si, which are noticeably lower than those for Mg$_{2}$Si. It is seen that the SERTA yields lower $\mu_{\mathrm{e}}$ values compared to IBTE. In contrast, the MRTA solution exhibits a different trend; at 300 K, it exceeds the IBTE value by almost 17\%, with the difference increasing monotonically to almost 30\% at 600 K and 51\% at 900 K. MRTA aligns better with IBTE result near room-temperature, while SERTA provides comparable results at higher temperatures. Previous studies on polar materials have also reported a similar trend of $\mu_{\mathrm{e}}$ values obtained using SERTA and IBTE \cite{ma2018first,ma2022first}. The SERTA, MRTA, and IBTE methods estimate $\mu_{\mathrm{e}}$ at 300 K to be $\sim$100, 197, and 163 {\cvs}, respectively. These values reduce to $\sim$11, 21, and 11 {\cvs} at 900 K, according to the respective methods.

\subsection{\label{sec:3d}Electron and phonon transport coefficients}
Various approaches have been developed to study charge carrier transport in TE materials within the Boltzmann equation formalism. The most common and accessible method for determining these properties is CRTA. The use of an energy-independent $\uptau$ in this approach is generally reasonable for good electrical conductors because the electron energy relaxation time $\uptau_{n\textbf{k}}$ does not vary significantly with $\varepsilon_{n\textbf{k}}$, except near $\varepsilon_{\mathrm{F}}$ \cite{singh1997calculated,ashcroft1976}. However, for insulators, employing CRTA to determine transport may be a poor approach, as $\uptau_{n\textbf{k}}$ is a critical factor that can affect the accuracy of the results and their subsequent comparison with experiment. 

In this work, we consider both aspects of $\uptau$ to analyze the electronic part of both TE silicides using different RTA approaches: (i) an energy-independent (constant) $\uptau$ within CRTA, and (ii) an energy-dependent lifetime $\uptau_{n\textbf{k}}$ based on EPI within SERTA and MRTA. In the CRTA method, the constant value of $\uptau$ is set to the value at 300 K, as obtained within the MRTA framework for each carrier concentration. The electrical components related to $z$T are derived using the electronic dispersion from PBE and the band gap from mBJ for both n-type compounds. This combination is employed because it provides TE properties, such as the Seebeck coefficient, that  match more closely with experimental observations \cite{sk2018exploring}. The gap values, taken from our previous study \cite{solet2024}, are 0.6 eV for Mg$_{2}$Si and 0.96 eV for Ca$_{2}$Si.

\subsubsection{\label{sec:3d1}Seebeck coefficient}

\begin{figure*}[ht]
\includegraphics[width=16.3cm, height=8.0cm]{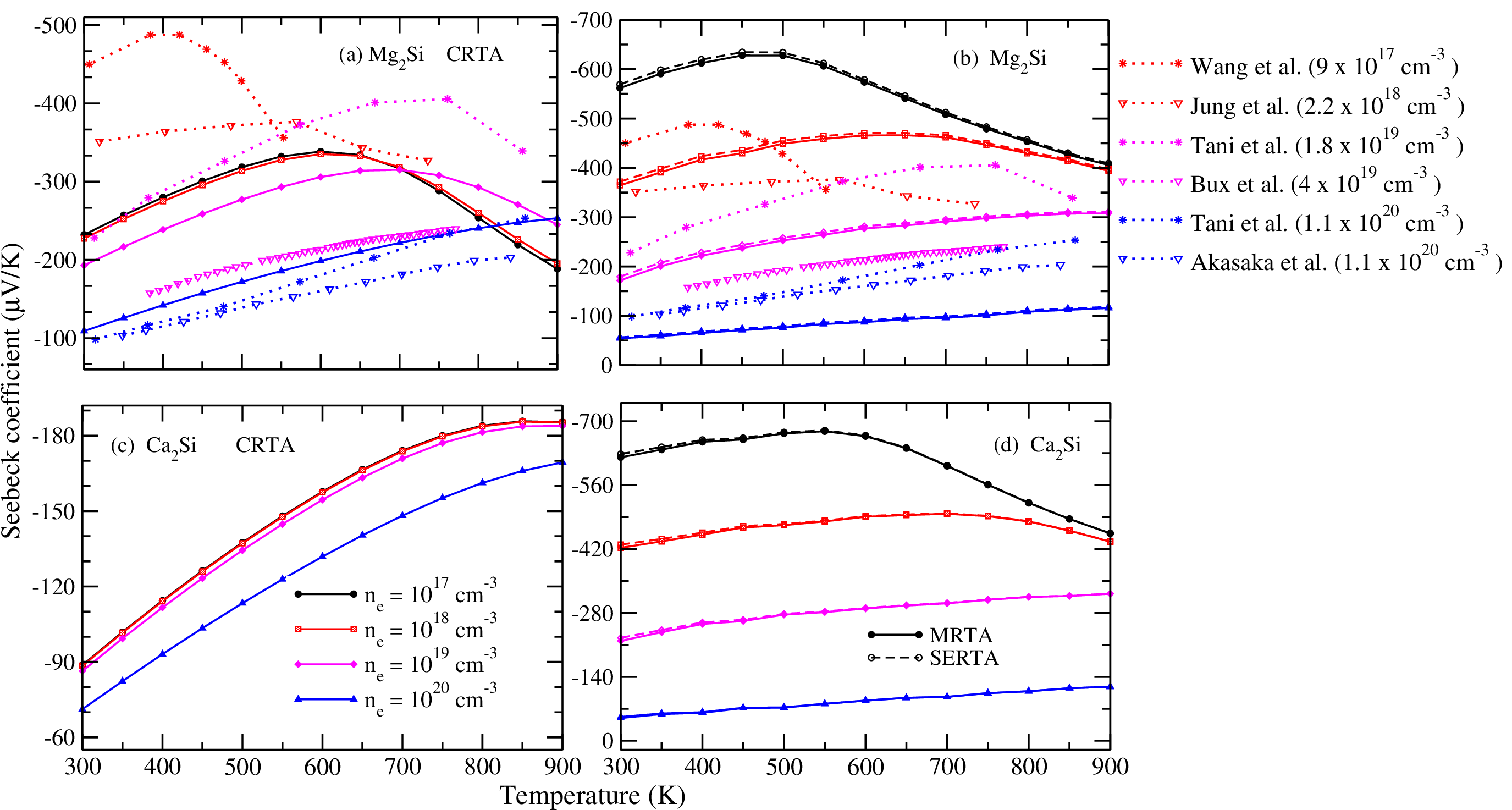} 
\caption{Seebeck coefficient as a function of temperature at different electron doping densities (n$_{\mathrm{e}}$) for $n$-type Mg$_{2}$Si and Ca$_{2}$Si, calculated using [(a), (c)] the CRTA and [(b), (d)] the MRTA (solid lines) and SERTA (dashed lines). Experimental data (dotted lines) for Mg$_{2}$Si in [(a), (b)] for different n$_{\mathrm{e}}$ values are shown: 9 $\times$ $10^{17}$ {\cm} (Wang $et$ $al.$ \cite{wang2012theoretical}), 2.2 $\times$ $10^{18}$ {\cm} (Jung $et$ $al.$ \cite{jung2011solid}), 1.8 $\times$ $10^{19}$ and 1.1 $\times$ $10^{20}$ {\cm} (Tani $et$ $al.$ \cite{tani2005thermoelectric}), 4 $\times$ $10^{19}$ {\cm} (Bux $et$ $al.$ \cite{bux2011mechanochemical}), and 1.1 $\times$ $10^{20}$ {\cm} (Akasaka $et$ $al.$ \cite{akasaka2008thermoelectric}).}  
\label{fig:sbk}
\end{figure*}

In this section, we have plotted Seebeck coefficient (S) for both $n$-type materials over a T range of 300-900 K in a wide range of electron concentrations (n$_{\mathrm{e}}$) from $10^{17}$ to $10^{20}$ {\cm} in Fig.~\ref{fig:sbk}. First we can observe in Figs.~\hyperref[fig:sbk]{\ref{fig:sbk}(b)} and ~\hyperref[fig:sbk]{\ref{fig:sbk}(d)} that the SERTA and MRTA methods yield nearly similar values of S across the entire T range. This indicates that forward and backward scattering are approximately equal when defining S using these two RTAs. Figure~\hyperref[fig:sbk]{\ref{fig:sbk}(a)} shows that the magnitude of Seebeck coefficient ($|\mathrm{S}|$) using CRTA decreases, especially up to $\sim$650 K with increasing n$_{\mathrm{e}}$ for Mg$_{2}$Si. One can also observe that difference between calculated CRTA result and experimental values decreases with increasing carrier concentration from $10^{17}$ to $10^{20}$ {\cm}. This means that one can reproduce closer results with experiments using CRTA at contractions around 10$^{19}$-10$^{20}$ {\cm}. This can be observe in the figure where the calculated S at $10^{20}$ {\cm} matches very well with experimental data obtained by Tani $et$ $al.$ at 1.1 $\times$ $10^{20}$ {\cm} \cite{tani2005thermoelectric}. However, at same concentration, our result has less accuracy, particularly for higher temperatures, with experimental data reported by Akasaka $et$ $al.$ \cite{akasaka2008thermoelectric} as compared to Tani $et$ $al.$, possibly can be due to differences in sample fabrication methods used in both study. Also the obtained S at $10^{19}$ {\cm} has almost similar trend with experiment performed by Tani $et$ $al.$ at 1.8 $\times$ $10^{19}$ {\cm} \cite{tani2005thermoelectric}.  

Figure~\hyperref[fig:sbk]{\ref{fig:sbk}(b)} compares the results when EPI is included over CRTA for Mg$_{2}$Si with experimental data. It is observed that, the values of S decrease with increasing concentrations from $10^{17}$ to $10^{20}$ {\cm}. This trend aligns with the typical behavior of most semiconductors, where higher carrier concentrations result in a reduction of S. Notably, the inclusion of EPI significantly affects the values of S compared to CRTA result. Due to this, S matches much better with experimental data reported by Wang $et$ $al.$ \cite{wang2012theoretical} at 9 $\times$ $10^{17}$ {\cm} and by Jung $et$ $al.$ \cite{jung2011solid} at 2.2 $\times$ $10^{18}$ {\cm}, particularly around room-temperature for lower n$_{\mathrm{e}}$ values ($10^{17}$-$10^{18}$ {\cm}). Thus, one should not ignore scattering mechanisms, although they are directly related to carrier lifetime, for the accurate estimation of S. The S value at $10^{17}$ {\cm} shows an increasing trend up to 500 K, followed by a decrease at higher T due to the thermal excitation of holes. For n$_{\mathrm{e}}$ = $10^{19}$ {\cm}, the value of S shows a linear increase with T, a behavior consistent with the experimental observations by Bux $et$ $al.$ \cite{bux2011mechanochemical} at 4 $\times$ $10^{19}$ {\cm}. This linear behavior is characteristic of heavily doped semiconductors. The strong effect of EPI on concentration for $10^{20}$ {\cm} is observed where the calculated S is significantly reduced compared to CRTA results and underestimates the experimental data, largely at higher temperatures \cite{tani2005thermoelectric,akasaka2008thermoelectric}. Discrepancies between theoretical and experimental S values may stem from defects, disorders, or off-stoichiometry in real samples, as calculations assume ideal stoichiometric compounds. Experimental challenges like contact resistance or precisely maintaining the temperature gradient, along with computational limitations such as the use of rigid band approximation for doping and neglect of band structure renormalization in transport calculations, may also contribute. 

Figure~\hyperref[fig:sbk]{\ref{fig:sbk}(c)} for Ca$_{2}$Si shows that $|\mathrm{S}|$ obtained from CRTA increases with T at the same n$_{\mathrm{e}}$ and decreases with n$_{\mathrm{e}}$ at the same T, a trend similar to  observed in most TE semiconductors. But the change is almost negligible for first two concentrations and then $|\mathrm{S}|$ shows slight changes with n$_{\mathrm{e}}$ from $10^{19}$ to $10^{20}$ {\cm}. At 300 K, the range of obtained $|\mathrm{S}|$ is $\sim$70-89 {\sbku} for decreasing concentrations from $10^{20}$ to $10^{17}$ {\cm}, which further increases to $\sim$170-185 {\sbku} at 900 K. In Fig.~\hyperref[fig:sbk]{\ref{fig:sbk}(d)}, incorporating energy-dependent $\uptau$ via EPI significantly increases $|\mathrm{S}|$ for low concentrations ($10^{17}$-$10^{19}$ {\cm}) over the studied T range but decreases $|\mathrm{S}|$ at n$_{\mathrm{e}}$ of $10^{20}$ {\cm} compared to the CRTA calculation. The highest obtained $|\mathrm{S}|$ is $\sim$678 {\sbku} at 500 K for n$_{\mathrm{e}}$ of $10^{17}$ {\cm}, which decreases to $\sim$500 {\sbku} at 700 K for a concentration of $10^{18}$ {\cm}. For n$_{\mathrm{e}}$ ranging from $10^{19}$ to $10^{20}$ {\cm}, maximum $|\mathrm{S}|$ decreases from $\sim$321 to 118 {\sbku} at 900 K. The calculated room-temperature $|\mathrm{S}|$ of $\sim$220 {\sbku} using EPI for a n$_{\mathrm{e}}$ of $10^{19}$ {\cm} is higher than that of Mg$_{2}$Si and other TEs such as PbTe at the same concentration \cite{heremans2008enhancement}. Generally, the typical S values for high-$z$T TE materials varies between $\sim$200-300 {\sbku}, these obtained results for S hint the possibility of achieving a high $z$T in both $n$-type materials. 

\subsubsection{\label{sec:3d2}Electrical conductivity}

\begin{figure*}[ht]
\includegraphics[width=15.8cm, height=6.5cm]{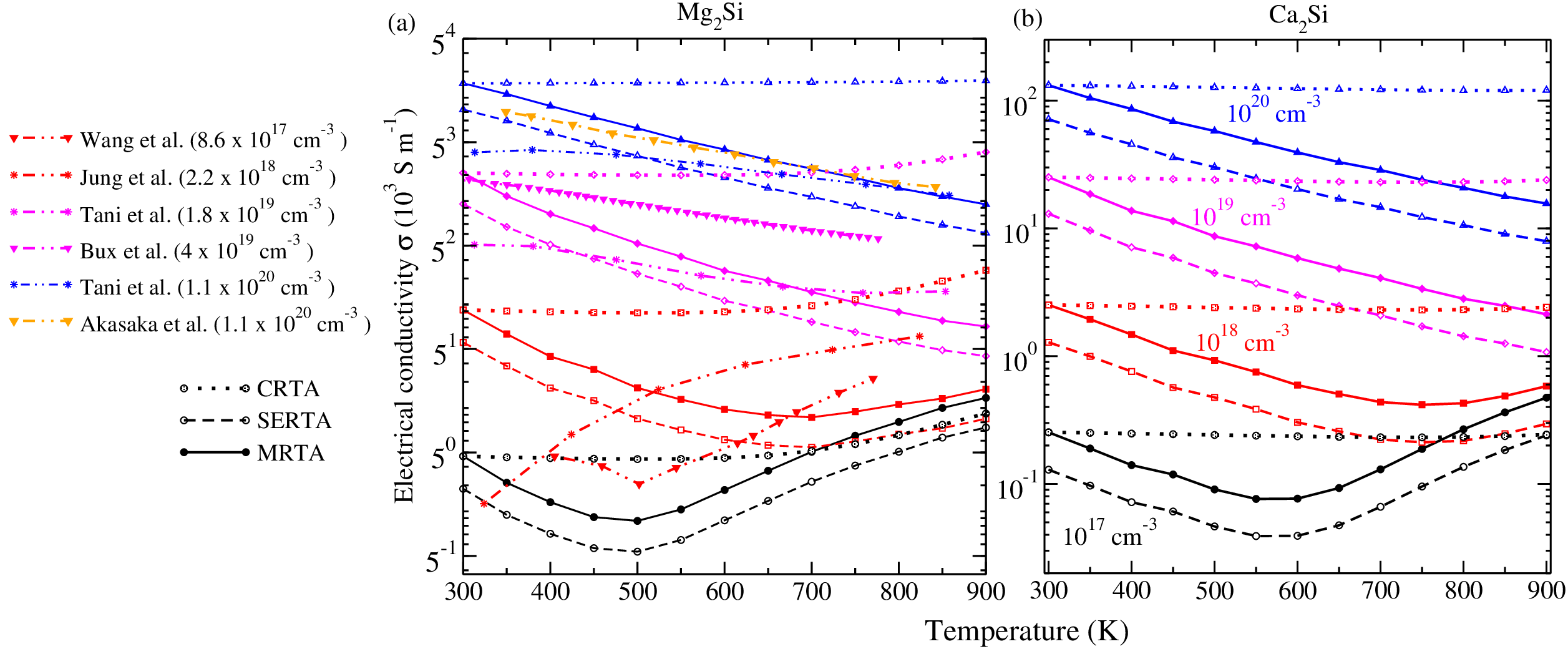} 
\caption{Calculated temperature-dependent electrical conductivity ($\sigma$) at similar doping densities, using all three RTAs and the same experimental references for $\sigma$ data (shown as dashed-dotted-dotted lines) as in Fig.~\hyperref[fig:sbk]{\ref{fig:sbk}}.}  
\label{fig:sigma}
\end{figure*} 

Figure~\ref{fig:sigma} compares the electrical conductivity ($\sigma$) obtained within various RTAs in both $n$-type materials. In the CRTA approach, $\uptau$ value is treated as constant, so the calculated $\sigma$ is hardly improved with changes in T at a particular n$_{\mathrm{e}}$. This behavior is evident for both compounds where $\sigma$ values obtained using CRTA show very little variation as T increases. The SERTA method provides lower values than MRTA, consistent with the findings for $\mu_{\mathrm{e}}$ depicted in Fig.~\ref{fig:mob}. However, the use of energy-dependent $\uptau$ from EPI within the SERTA and MRTA methods significantly affects $\sigma$ results across all concentrations, especially for $10^{18}$-$10^{20}$ {\cm} in both materials. This inclusion thus reflects results that more closely match experimental data compared to CRTA, as illustrated in Fig.~\hyperref[fig:sigma]{\ref{fig:sigma}(a)} for Mg$_{2}$Si. For example, $\sigma$ obtained using SERTA at $10^{17}$ {\cm} is lower than that predicted by CRTA, while a same trend for MRTA is observed up to $\sim$700 K after which it surpasses the CRTA predictions. Both SERTA and MRTA exhibit trends consistent with experimental data reported by Wang $et$ $al.$ \cite{wang2012theoretical} for a concentration of 8.6 $\times$ $10^{17}$ {\cm}. For higher concentrations ($10^{19}$-$10^{20}$ {\cm}), the inclusion of EPI into CRTA provides an estimate that decreases almost linearly with T, in contrast to the almost unchanged behavior with T observed using CRTA. This indicates that $\uptau$ plays a particularly important role at higher carrier concentrations in reproducing results closer to experimental observations, as seen in Fig.~\hyperref[fig:sigma]{\ref{fig:sigma}(a)}. The computed results for higher concentrations are also significantly improved compared to those reported by Fan $et$ $al.$ \cite{fan2018first}. Aside from the reasons for the mismatch between the calculated and experimental results discussed in the previous section, $\uptau$ can also be influenced by other scattering mechanisms, such as electron-electron interactions \cite{bernardi2014ab}, which are not considered here for calculating $\sigma$. 

Figure~\hyperref[fig:sigma]{\ref{fig:sigma}(b)} for Ca$_{2}$Si shows that $\sigma$ decreases with increasing T, reaching its lowest value for the first two n$_{\mathrm{e}}$ values, before increasing again. For n$_{\mathrm{e}}$ values of $10^{19}$ and $10^{20}$ {\cm}, $\sigma$ decreases almost linearly with T over the considered region. A similar trend for $\sigma$ with n$_{\mathrm{e}}$ is also observed in Mg$_{2}$Si. Notably, for last three n$_{\mathrm{e}}$ values, the inclusion of EPI via both SERTA and MRTA reduces $\sigma$ compared to CRTA at the respective concentrations in Ca$_{2}$Si. This means that dominance of $\uptau$ in obtaining $\sigma$ increases as T increases, obvious due to increased carrier scattering with phonons. One can observe, in both materials, that the $\sigma$ obtained using SERTA and MRTA for first two n$_{\mathrm{e}}$ values approaches the same value. The causes may be that the electron density for both doping concentrations becomes almost similar at higher temperatures. Overall, the $\sigma$ is found to be higher for Mg$_{2}$Si than for Ca$_{2}$Si using all three studied methods.

\subsubsection{\label{sec:3d3}Thermal conductivity}
\begin{figure}
\includegraphics[width=8.0cm, height=8.8cm]{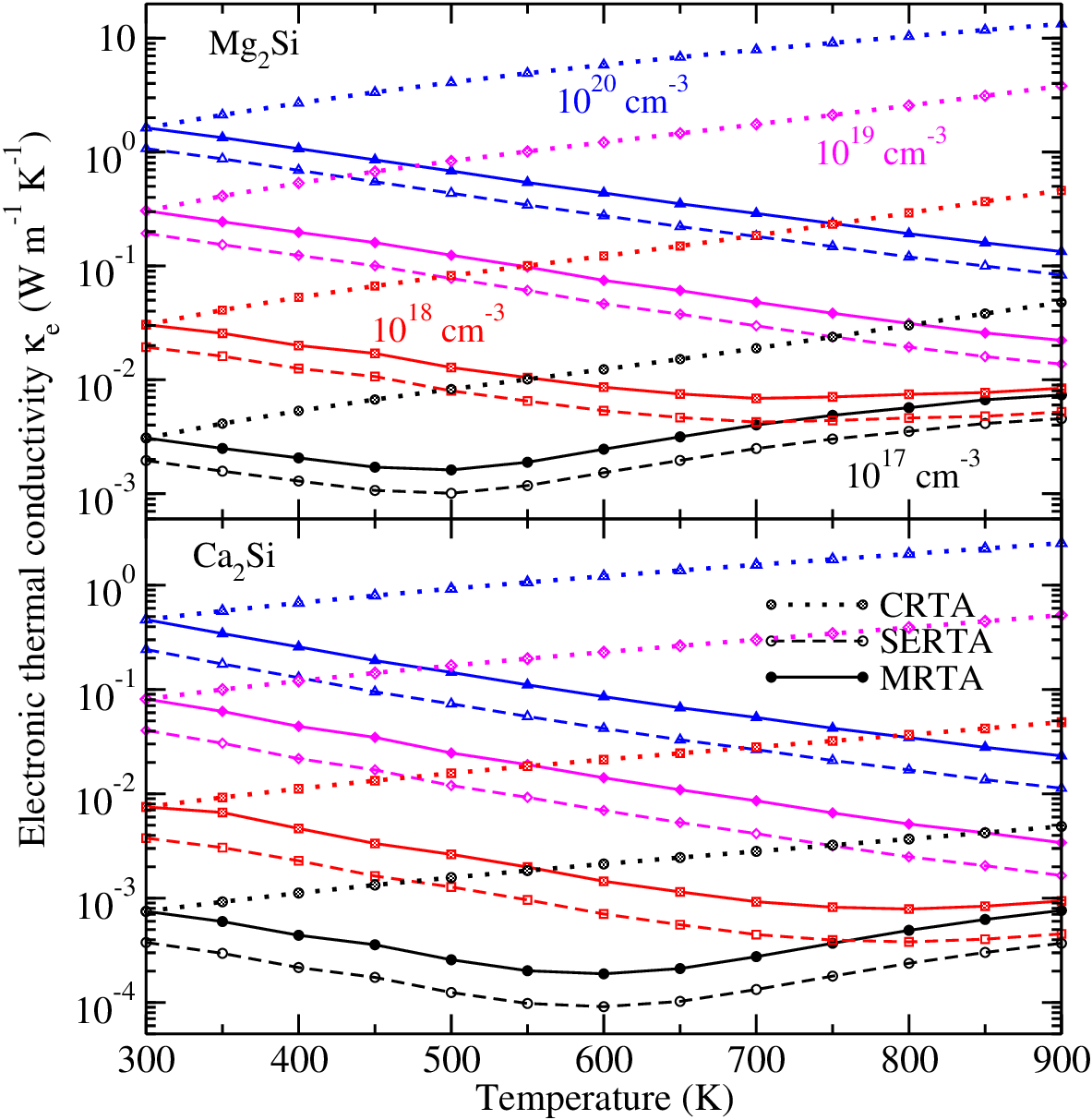} 
\caption{Electronic part of the thermal conductivity ($\kappa_{\mathrm{e}}$) as a function of temperature at different electron doping densities for $n$-type Mg$_{2}$Si and Ca$_{2}$Si, calculated using all three RTAs.}  
\label{fig:ke}
\end{figure}

We now turn our attention to another crucial facet of its TE performance: thermal conductivity. To gain insight into this transport property, we first calculate thermal conductivity due to the carrier transport ($\kappa_{\mathrm{e}}$) both without and with inclusion of EPI picture into CRTA, as shown in Fig.~\ref{fig:ke}. In this figure, $\kappa_{\mathrm{e}}$ from CRTA increases with increasing T for all values of n$_{\mathrm{e}}$ used. The behavior of $\kappa_{\mathrm{e}}$ with T is similar to what is observed for $\sigma$, as both are directly related to the Wiedemann-Franz law. As with $\sigma$, the use of energy-dependent $\uptau$ in CRTA alters the T-dependent behavior of $\kappa_{\mathrm{e}}$, now showing a decreasing trend with T. The $\kappa_{\mathrm{e}}$ increases with increasing concentrations, a trend similar to what has been observed in experimental studies. At higher concentrations ($10^{19}$-$10^{20}$ {\cm}), the decreasing trend of $\kappa_{\mathrm{e}}$ is consistent with the experimentally observed data for Mg$_{2}$Si at the same concentration order \cite{bux2011mechanochemical}. In general, in nonmetallic systems, phonons are the primary contributors to the total thermal conductivity, meaning they primarily determine the thermal behavior in TE materials.

\begin{figure}
\includegraphics[width=7.3cm, height=8.7cm]{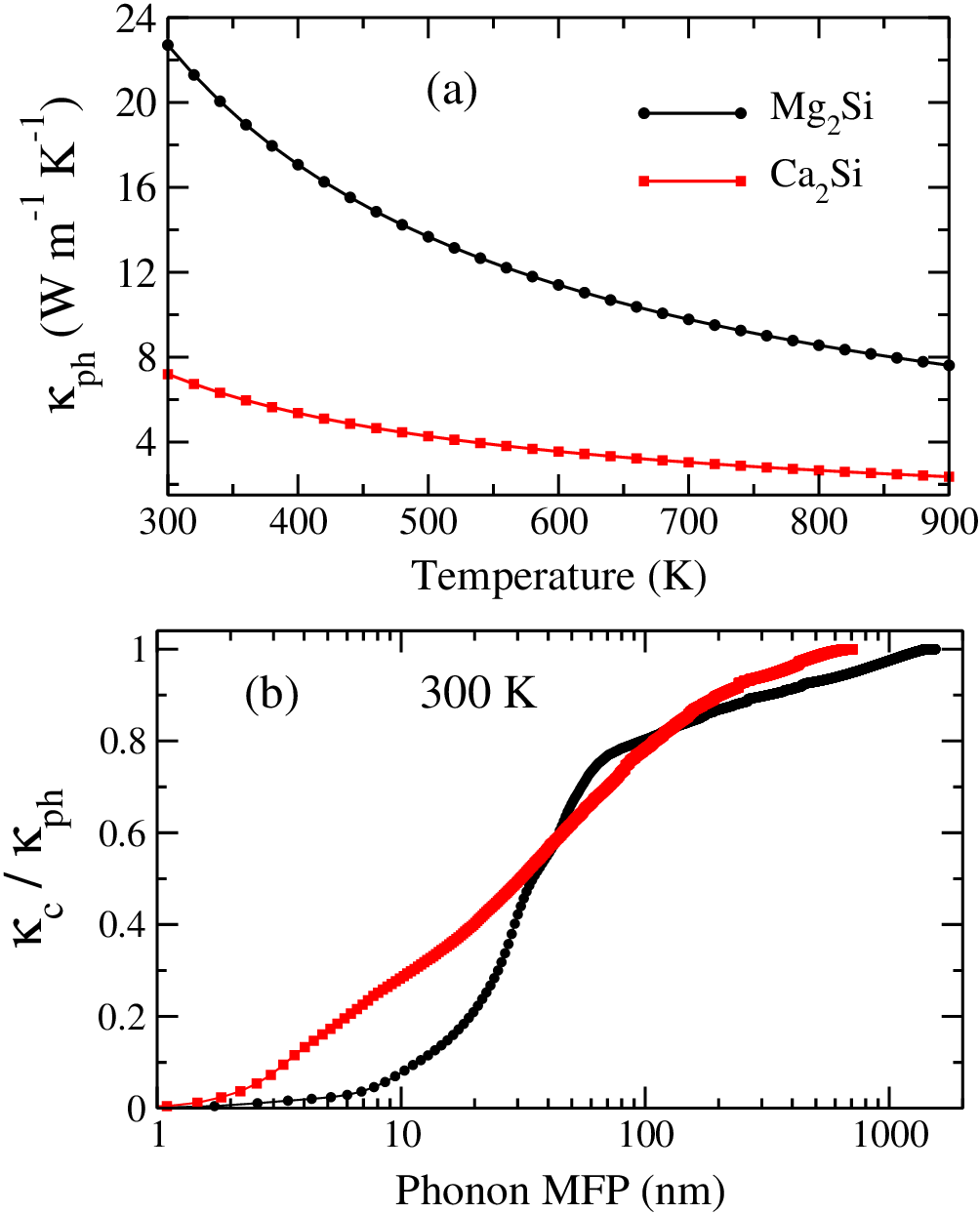} 
\caption{(a) Calculated lattice thermal conductivity ($\kappa_{\mathrm{ph}}$) from phonon-phonon interaction as a function of temperature. (b) Normalized cumulative lattice thermal conductivity ($\kappa_{c}$/$\kappa_{\mathrm{ph}}$) at room temperature as a function of phonon mean-free-path (MFP).} 
\label{fig:kph}
\end{figure}

\begin{figure*}
\includegraphics[width=14.2cm, height=7.1cm]{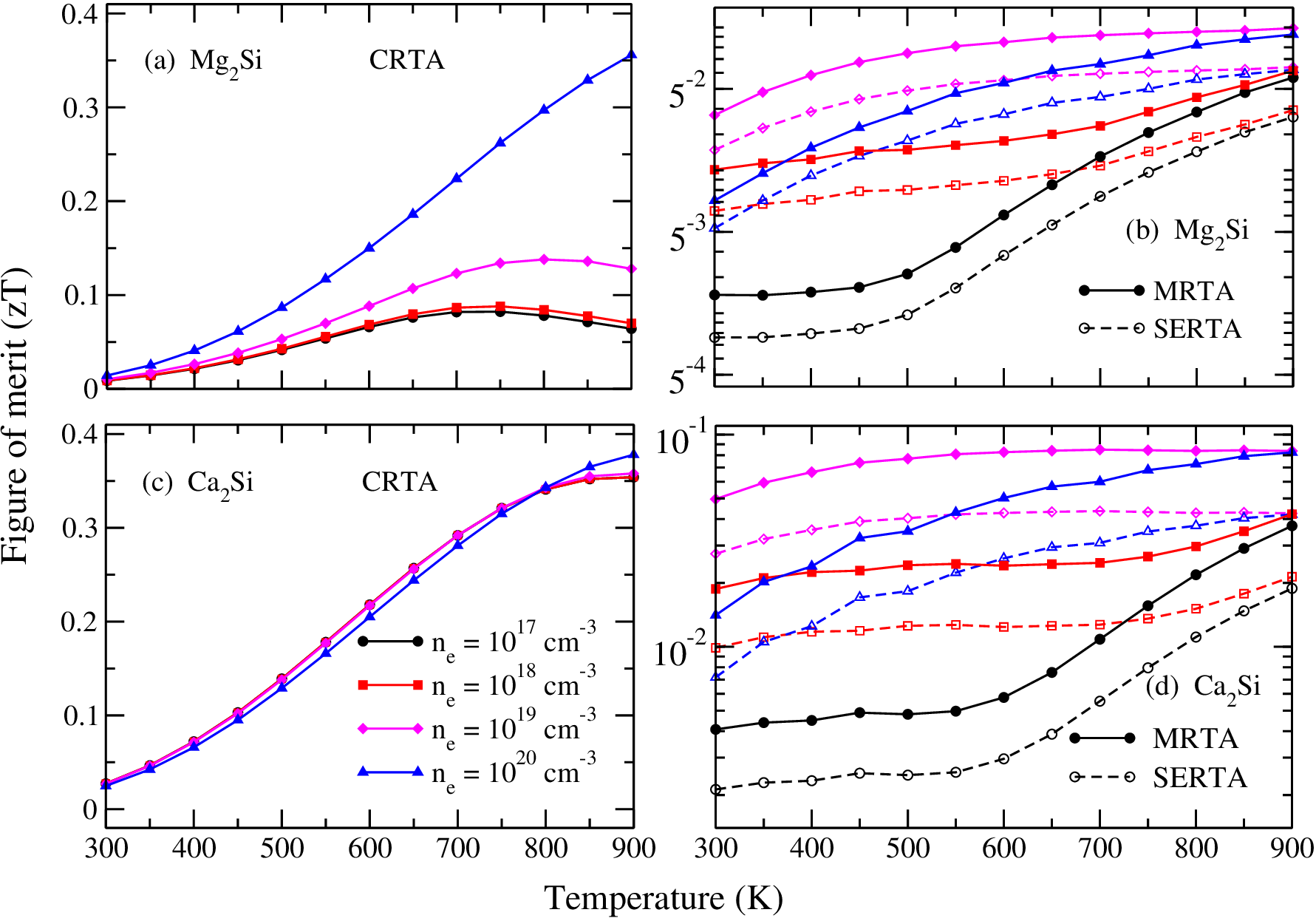} 
\caption{For $n$-type Mg$_{2}$Si and Ca$_{2}$Si, electron concentration (n$_{\mathrm{e}}$) dependence of figure of merit ($z$T) as a function of temperature: CRTA in panels (a) and (c), and MRTA (solid line) and SERTA (dashed line) in panels (b) and (d).} 
\label{fig:zt}
\end{figure*}

Thus, to understand the thermal behavior of heat energy transported by phonons, the $\kappa_{\mathrm{ph}}$ is estimated using the PPI formalism over the temperature range of 300–900 K. Figure~\hyperref[fig:kph]{\ref{fig:kph}(a)} shows that $\kappa_{\mathrm{ph}}$ continues to decrease with increasing T for both materials, which is obvious because PPI becomes more frequent and energetic as T raises. At all studied T values, $\kappa_{\mathrm{ph}}$ for Ca$_{2}$Si is lower than that for Mg$_{2}$Si, and one possible reason for this behaviour is the presence of heavier calcium atoms compared to the lighter magnesium atoms in the studied silicides. For example, the obtained $\kappa_{\mathrm{ph}}$ at 300 K is $\sim$22.7 (7.2) W m$^{-1}$ K$^{-1}$ for Mg$_{2}$Si (Ca$_{2}$Si), which further reduces to $\sim$7.6 (2.4) at 900 K. In summary, above 300 K, both $\kappa_{\mathrm{e}}$ and $\kappa_{\mathrm{ph}}$ in studied semiconductors decrease with increasing T due to enhanced scattering effects$-$e-p scattering for the electronic component and p-p scattering for the lattice component. This outcome clearly emphasizes the importance of both including scattering mechanisms, as they markedly reduce $\kappa$, particularly for higher n$_{\mathrm{e}}$ and T, in contrast to the values derived from CRTA. Such a reduction suggests a substantial benefit in employing both materials for high-T TE applications, where lower $\kappa$ can enhance performance.

\subsubsection{\label{sec:3d4}Figure of merit ($z$T)}
We now have all the necessary parameters to access the performance of both $n$-type TE silicides, primarily determined by $z$T shown in Fig.~\ref{fig:zt}. For Mg$_{2}$Si using CRTA, the $z$T value increases with T for all n$_{\mathrm{e}}$ values; however at lower concentrations ($10^{17}$ to $10^{19}$ {\cm}), it remains nearly constant in the high-T region. The highest $z$T range for these three concentrations is $\sim$0.08-0.14 at T$\approx$750 K. The nearly linear increase of $z$T with T for n$_{\mathrm{e}}$ = $10^{20}$ {\cm} reaches a maximum of $\sim$0.35 at 900 K, which is considerably higher than at lower concentrations. A surprising feature of $z$T for n$_{\mathrm{e}}$ appears in Ca$_{2}$Si using CRTA, as shown in Fig.~\hyperref[fig:zt]{\ref{fig:zt}(c)}. In this figure, increasing the concentration from lower ($10^{17}$ {\cm}) to higher ($10^{20}$ {\cm}) values does not affect $z$T in significant way within the studied T range, where the highest obtained values are in range of $\sim$0.35-0.38 at 900 K.

Figures~\hyperref[fig:zt]{\ref{fig:zt}(b)} and ~\hyperref[fig:zt]{\ref{fig:zt}(d)} show the $z$T values when EPI is incorporated into the CRTA for both materials. This presence significantly alters the behavior of $z$T across both n$_{\mathrm{e}}$ and T in both materials. The effect of EPI on $z$T is stronger in Ca$_{2}$Si than in Mg$_{2}$Si, same as observed in previously discussed electronic part. In both materials, the $z$T obtained from MRTA for each n$_{\mathrm{e}}$ and T is consistently higher than the value obtained from SERTA. This indicates that we cannot ignore back-scattering electron processes, as they essentially improve $z$T. In both materials, optimal $z$T at all T is observed for optimal n$_{\mathrm{e}}$ value of $10^{19}$ {\cm} using MRTA. At this n$_{\mathrm{e}}$, the highest $z$T is $\sim$0.03 (0.08) for Mg$_{2}$Si at 300 (900) K, and $\sim$0.05 (0.085) for Ca$_{2}$Si at 300 (700) K. But at T$\approx$900 K, last two concentration using MRTA provide almost similar $z$T. A donor concentration of $10^{19}$ electrons/cm$^{3}$ corresponds to electron doping of $\sim$0.00065 and 0.00092 electrons per unit cell in Mg$_{2}$Si and Ca$_{2}$Si, respectively. Achieving this doping concentration requires less than 1\% doping, which is experimentally feasible and desirable to dope both materials with donors in the range of $10^{19}$-$10^{20}$ {\cm} to optimize $z$T at high T region.  

\subsubsection{\label{sec:3d5}Strategies for enhancing $z$T}

It is important to note that the predicted $z$T values for both materials at n$_{\mathrm{e}}$ and T, based solely on EPI and PPI mechanisms, could represent the lowest possible values. In real TE materials, additional scattering channels such as impurity, boundary, or defect scattering are often present for both particles, and these can significantly reduce $\kappa$, especially $\kappa_{\mathrm{ph}}$. This leads to a significant enhancement in $z$T due to a more favorable balance between the numerator part (S$^{2}$$\sigma$) and $\kappa_{\mathrm{e}}$ in denominator part. For instance, in both silicides at optimal carrier concentration of $10^{19}$ {\cm} and 300 K using the MRTA approach, more than 98\% of the heat is transported by phonons, so there would be a possibility in the reduction of this percentage. One common technique to achieve this reduction is nanostructuring, which involves introducing grain boundaries or nanoscale precipitates \cite{minnich2009modeling,dresselhaus2007new,qiu2015first}. This method reduces the $\kappa_{\mathrm{ph}}$ by increasing phonon-boundary scattering, which is influenced by grain size \cite{zhou2010semiclassical,satyala2012detrimental,li2012thermal}. The effectiveness of nanostructuring for a specific material and the optimal length scale of nanostructures depend on detailed knowledge of the mean-free-path (MFP) distributions of electrons and phonons in that material. In typical TE materials, phonon MFPs \cite{tian2013heat} are much longer than electron MFPs \cite{qiu2015first}, so nanostructures with intermediate sizes can effectively impede heat conduction while leaving electron flow mostly unaffected. This approach can notably lower $\kappa_{\mathrm{ph}}$ in nanostructured materials, especially when the characteristic size of the nanoinclusions is less than the phonon MFP, which is generally in the range of tens to hundreds of nanometers (nm) in most semiconductors \cite{qiu2015first,chen2004encyclopedia}.

\begin{figure}[ht]
\includegraphics[width=7.3cm, height=8.4cm]{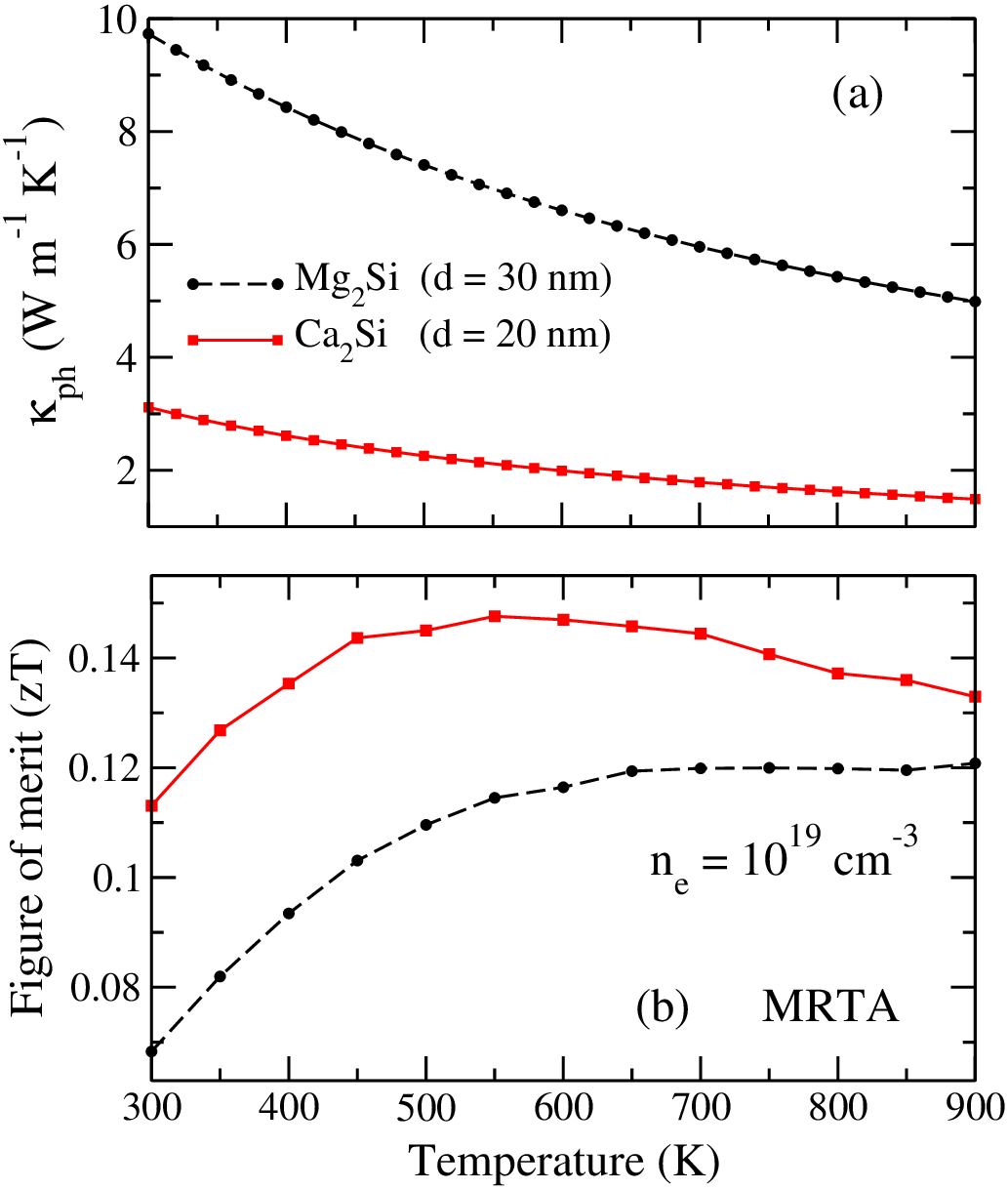} 
\caption{Temperature dependence of (a) $\kappa_{\mathrm{ph}}$ by including also phonon boundary scattering (d = 30 nm for Mg$_{2}$Si and 20 nm for Ca$_{2}$Si) and (b) corresponding $z$T at n$_{\mathrm{e}}$ = $10^{19}$ {\cm} using MRTA.} 
\label{fig:ztb}
\end{figure}

To understand how heat energy transport due to phonons is affected by nanostructuring, it is helpful to create an accumulation plot of $\kappa_{\mathrm{ph}}$ with respect to phonon MFPs. The contribution to thermal conductivity from phonons with MFPs ($\textbf{I}_{\lambda}$) up to $l$ can be estimated by summing over contributions from all phonon modes with MFPs shorter or equal to $l$, given by $\kappa_{\mathrm{c}}(l)$ = $\int^{l}_{0}(1/\mathrm{N})\sum_{\lambda}\kappa_{\lambda}\delta(l_{\lambda}-l^{\prime})dl^{\prime}$, where $l_{\lambda} = |\textbf{I}_{\lambda}|= \textbf{v}_{\lambda}\uptau_{\lambda}$ and $\kappa_{\lambda}$ = $(1/\mathrm{V}_{0})\mathrm{C}_{\lambda}\textbf{v}_{\lambda}\otimes\textbf{I}_{\lambda}$ represents the contribution to $\kappa_{\mathrm{ph}}$ in mode $\lambda$ under the SMRTA approach \cite{togo2023first}, as discussed in Sec.~\ref{sec:2c}. A clear view of MFP distributions is given by the normalized \textquotedblleft cumulative lattice thermal conductivity \textquotedblright, $\kappa_{\mathrm{c}}$/$\kappa_{\mathrm{ph}}$. Figure~\hyperref[fig:kph]{\ref{fig:kph}(b)} includes the plot of $\kappa_{\mathrm{c}}$/$\kappa_{\mathrm{ph}}$ for both silicides at 300 K normalized by the room-temperature bulk $\kappa_{\mathrm{ph}}$ value. It reveals that phonons at room-temperature carry more than 40\% of heat for MFPs below 30 nm for Mg$_{2}$Si and 20 nm for Ca$_{2}$Si. However $\kappa_{\mathrm{ph}}$ has contributions from phonons with MFPs up to 1 $\mu$m, as observed in Fig.~\hyperref[fig:kph]{\ref{fig:kph}(b)}. To effectively increase $z$T through nanostructuring, the nanostructure size should be smaller than the phonon MFPs but large than the electron MFPs, so that phonons are more likely to scattered than electrons. Previous studies on Mg$_{2}$Si \cite{satyala2012effect,satyala2012detrimental} as well as other semiconductors like Si \cite{qiu2015first} indicate that the majority part of electrical conductivity contributions comes from electrons with MFPs less than 10 nm. In other words, choosing a grain size above 10 nm does not have a significant impact on electrical transport but can effectively reduce $\kappa_{\mathrm{ph}}$. Therefore, we can choose nanostructure with sizes between 10 nm and 1 $\mu$m.

Based on this approach and Fig.~\hyperref[fig:kph]{\ref{fig:kph}(b)}, nanostructure with size d = 30 (20) nm in Mg$_{2}$Si (Ca$_{2}$Si) can be selected to maximize phonon scattering while preserving most electron transport, in order to minimize $\kappa_{\mathrm{ph}}$. We then re-compute the $\kappa_{\mathrm{ph}}$ by incorporating also the boundary scattering rates (v$_{\mathrm{g}}$/d) where v$_{\mathrm{g}}$ is the group velocity and d is the boundary MFP. This recalculation is shown in Fig.~\hyperref[fig:ztb]{\ref{fig:ztb}(a)} for the chosen d values in both materials. Boundary scattering leads to a significant reduction $\kappa_{\mathrm{ph}}$, which is $\sim$9.7 (3.1) W m$^{-1}$ K$^{-1}$ in Mg$_{2}$Si (Ca$_{2}$Si) at 300 K$-$ a decrease of almost 55-60\% compared to the bulk values in both materials. The value at high T (900 K) is reduced by almost 35\% compared to bulk values. This slower reduction at elevated temperatures may be attributed to the reduced dominance of phonon-boundary scattering as T increases. The obtained room-temperature value for Mg$_{2}$Si falls within the experimentally observed range of $\sim$7.8-10.5 W m$^{-1}$ K$^{-1}$ \cite{labotz1963thermal,tani2005thermoelectric,rowe2018thermoelectrics}. The accuracy of the calculated $\kappa_{\mathrm{ph}}$ may be enhanced by using a supercell size beyond 2 $\times$ 2 $\times$ 2, as well as increasing the cutoff pair distance used in the present calculations. Using the $\kappa_{\mathrm{ph}}$ of nanostructured Mg$_{2}$Si and Ca$_{2}$Si, we again compute the $z$T and plot it in Fig.~\hyperref[fig:ztb]{\ref{fig:ztb}(b)} at the optimal electron concentration of $10^{19}$ {\cm} using MRTA. At 300 K, the use of nanograins results in more than a twofold enhancement in their $z$T compared to the bulk level. While the optimal $z$T is predicted as $\sim$0.12 for Mg$_{2}$Si at higher temperatures and $\sim$0.15 at mid T range (around 550 K) for Ca$_{2}$Si, exceeding their respective optimal bulk values by more than 1.5 times.

\begin{figure}[ht]
\includegraphics[width=7.3cm, height=8.4cm]{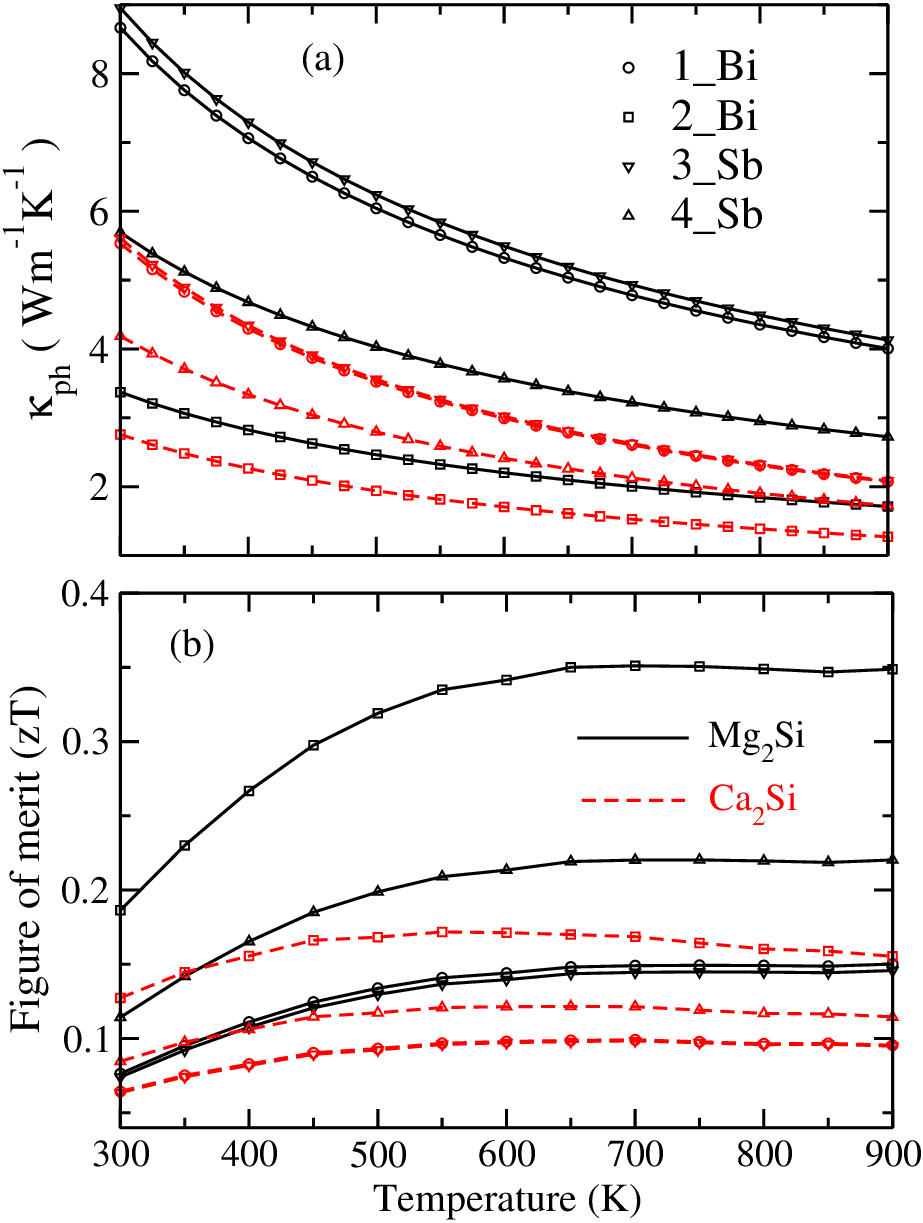} 
\caption{Temperature dependence of (a) $\kappa_{\mathrm{ph}}$ and (b) corresponding $z$T at $10^{19}$ {\cm} using MRTA for four n-type doped samples. Here 1\_Bi, 2\_Bi, 3\_Sb, and 4\_Sb correspond to 0.15\% Bi doping, 2\% Bi doping, 0.5\% Sb doping, and 2\% Sb doping, respectively, in both Mg$_{2}$Si (black) and Ca$_{2}$Si (red).} 
\label{fig:ztd}
\end{figure}

In experimental studies of Mg$_{2}$Si \cite{tani2005thermoelectric,bux2011mechanochemical,akasaka2008thermoelectric,tani2007thermoelectric_sb}, it has been observed that doping is essential for enhancing the $z$T. To fully realize the potential of Mg$_{2}$Si and Ca$_{2}$Si for TE applications, it is important to explore their solid solutions. Mg$_{2}$Si exhibits n-type conduction when doped with atoms such as Sb and Bi \cite{tani2005thermoelectric,tani2007thermoelectric_sb,bux2011mechanochemical}. These studies have also revealed that the highest $z$T values for Mg$_{2}$Si are achieved with specific doping concentrations: 0.15\% \cite{bux2011mechanochemical} and 2\% \cite{tani2005thermoelectric} Bi, as well as 0.5\% and 2\% Sb \cite{tani2007thermoelectric_sb}, corresponding to electron concentrations in the range of $10^{19}$-$10^{20}$ {\cm}. The obtained $\kappa_{\mathrm{ph}}$ values using these dopants are presented in Fig.~\hyperref[fig:ztd]{\ref{fig:ztd}(a)}. In Mg$_{2}$Si, compared to the pure compound, the $\kappa_{\mathrm{ph}}$ at 300 K decreases to almost 62\%, 85\%, 61\%, and 75\% when doped with 0.15\% and 2\% of Bi, and 0.5\% and 2\% of Sb, respectively. At 700 K, the $\kappa_{\mathrm{ph}}$ shows a smaller reduction, decreasing to almost 51\%, 80\%, 50\%, and 67\%, respectively. This indicates that with these small doping levels of Bi and Sb in pure Mg$_{2}$Si, the $\kappa_{\mathrm{ph}}$ value decreases by more than half. The obtained values for 0.15\% Bi and 0.5\% Sb doping are slightly higher than the experimental values reported in Refs. \cite{bux2011mechanochemical,tani2007thermoelectric_sb}, while the value for the 2\% Bi-doped sample is slightly lower, and the value for the 2\% Sb-doped sample is almost comparable to the experimental results \cite{tani2005thermoelectric,tani2007thermoelectric_sb}. Thus, the reasonable accuracy of the result can be obtained when we consider only simple mass difference scattering, and the observed discrepancies can be attributed to neglecting possible changes in inter-atomic force constants caused by doping. 

We further calculate the $z$T for these four Mg$_{2}$Si samples at optimal electron concentration within MRTA to evaluate the accuracy of the results compared to experimental data. As shown in Fig.~\hyperref[fig:ztd]{\ref{fig:ztd}(b)}, the highest $z$T values for the 2\% Bi- and Sb-doped samples are 0.35 and 0.22, respectively, which are about half of the highest experimental values. For the other two doped samples, about one-third of the experimental values is reproduced. This suggests that by neglecting factors due to doping, we have obtained the lower limit of $z$T for n-type doped Mg$_{2}$Si samples. Based on these results, we can further assess how much $z$T can be achieved for Ca$_{2}$Si at the same doping concentrations of both atoms using this approach.

Therefore, the $\kappa_{\mathrm{ph}}$ of Bi- and Sb-doped Ca$_{2}$Si samples is presented in Fig.~\hyperref[fig:ztd]{\ref{fig:ztd}(a)}. The values for all four samples are lower than that of the pure sample, with the 0.15\% of Bi-doped and 0.5\% Sb-doped samples showing almost similar values, meaning their $z$T values are also similar in our calculation. Among these samples, the 2\% Sb-doped compound has the lowest $\kappa_{\mathrm{ph}}$, suggesting it has the highest $z$T. The maximum $z$T for the 2\% Sb-doped Ca$_{2}$Si sample is found to be in the mid-temperature range, which is 0.17 for an electron concentration of $10^{19}$ {\cm} using MRTA. As a two-fold enhancement has been observed in experiments with Mg$_{2}$Si for this doping, a similar enhancement can be expected in Ca$_{2}$Si. Further, the reasonable $z$T can also be achieved by using nanostructuring in doped Ca$_{2}$Si sample. With this technique, we may expect a $z$T of around 0.4 in nanostructured 2\% Sb-doped Ca$_{2}$Si sample. Besides these studied techniques, many other factors, such as local strain within the lattice caused by defects \cite{yang2004strain}, also contribute to phonon scattering mechanisms. The effects of such factors can be analyzed in future $\kappa_{\mathrm{ph}}$ calculations.

Based on this analysis, we conclude that Ca$_{2}$Si can also be considered a potential material for TE and solar thermoelectric generators. Moreover, MFPs distributions for phonons offer valuable insights for experimentalists in engineering thermal transport in silicide nanostructures with optimal length scales. In our previous study \cite{solet2024}, Ca$_{2}$Si was identified as a highly efficient material for single-junction thin-film solar cells, achieving an optimal efficiency of $\sim$28.5\%. Thus, non-toxic, low cost Ca$_{2}$Si should be synthesized as it is the best dual-purpose material, functioning effectively as both solar cells and TE generators, offering a promising pathway for multifunctional energy solutions.

\section{Conclusions} \label{sec:concl}
In this work, we have performed first-principles calculations incorporating Boltzmann theory to compute electron transport using electron-phonon interaction (EPI) mechanism within different relaxation-time approximations (RTAs), and phonon-phonon interaction (PPI) to obtain the phonon transport of Mg$_{2}$Si and Ca$_{2}$Si compounds. The band gap renormalization, including both ZPR and T dependence upto 800 K, is analyzed. A detailed convergence study for room-temperature phonon-limited electron mobility $\mu_{\mathrm{e}}$ is performed using self-energy and momentum RTA (SERTA and MRTA), and the iterative solution of Boltzmann transport equation (IBTE), which are computed to be $\sim$351 (100), 573 (197), and 524 (163) \cvs, respectively, for Mg$_{2}$Si (Ca$_{2}$Si). Beyond 300 K, $\mu_{\mathrm{e}}$ decreases with T, reaching values at 900 K of $\sim$33 (11), 48 (21), and 28 (11) \cvs, respectively. Furthermore, the comparison between constant RTA (CRTA) and the EPI-based MRTA (or SERTA) methods helps in highlighting the suitability of the electron-phonon scattering mechanism for predicting the TE transport properties of silicides. Within single-mode RTA using PPI, the predicted $\kappa_{\mathrm{ph}}$ value at 300 K is $\sim$22.7 (7.2) W m$^{-1}$ K$^{-1}$ for Mg$_{2}$Si (Ca$_{2}$Si), reducing to $\sim$7.6 (2.4) W m$^{-1}$ K$^{-1}$ at 900 K. The maximum value of $z$T using CRTA is evaluated to be $\sim$0.35 (0.38), which reduces to $\sim$0.08 (0.085) under MRTA for an optimal carrier concentration of $10^{19}$ {\cm}. Next, we explore strategies to enhance $z$T by reducing $\kappa_{\mathrm{ph}}$ through nanostructuring and mass-difference scattering. The latter includes doping in Mg$_{2}$Si and Ca$_{2}$Si with Bi and Sb at specific concentrations. Our findings underscore the importance of EPI under different RTAs in accurately assessing the TE transport behavior of both silicides. Finally, we propose to synthesize Ca$_{2}$Si and characterize it experimentally to gain further insights into the current findings.

\vspace{0.0in}
\section*{Acknowledgement}
We acknowledge the computational support provided by the High-Performance Computing (HPC) PARAM Himalaya at the Indian Institute of Technology Mandi.
	
	

\bibliography{paper}

\end{document}